\documentclass[apj]{emulateapj}
\usepackage{amsmath}
\newcommand \p{\partial}

\def\dim#1{\mbox{\,#1}}

\begin{document}

  \title{Resolving Gas Dynamics in the Circumnuclear Region of a Disk Galaxy
  in a Cosmological Simulation}
  
  \author{Robyn Levine\altaffilmark{1,2,3}}
  \author{Nickolay Y. Gnedin\altaffilmark{3,4,5}}
  \author{Andrew J. S. Hamilton\altaffilmark{1,2}}
  \author{Andrey V. Kravtsov\altaffilmark{4,5,6}}
  \altaffiltext{1}{JILA, University of Colorado, Boulder, CO 80309;
  robyn.levine@colorado.edu}
  \altaffiltext{2}{Department of Astrophysical \& Planetary Sciences,
  University of Colorado, Boulder, CO 80309, USA}
  \altaffiltext{3}{Particle Astrophysics Center, Fermi National
  Accelerator Laboratory, Batavia, IL 60510, USA}
  \altaffiltext{4}{Kavli Institute for Cosmological Physics, The
  University of Chicago, Chicago, IL 60637, USA}
  \altaffiltext{5}{Department of Astronomy \& Astrophysics, The
  University of Chicago, Chicago, IL 60637, USA}
  \altaffiltext{6}{Enrico Fermi Institute, The University of Chicago,
  Chicago, IL 60637, USA}

  
  \begin{abstract}
    Using a hydrodynamic adaptive mesh refinement code, we simulate
    the growth and evolution of a galaxy, which could potentially host
    a supermassive black hole, within a cosmological volume. Reaching
    a dynamical range in excess of 10 million, the simulation follows
    the evolution of the gas structure from super-galactic scales all
    the way down to the outer edge of the accretion disk. Here, we
    focus on global instabilities in the self-gravitating, cold,
    turbulence-supported, molecular gas disk at the center of the
    model galaxy, which provide a natural mechanism for angular
    momentum transport down to sub-pc scales. The gas density profile
    follows a power-law $\propto r^{-8/3}$, consistent with an
    analytic description of turbulence in a quasi-stationary
    circumnuclear disk. We analyze the properties of the disk which
    contribute to the instabilities, and investigate the significance
    of instability for the galaxy's evolution and the growth of a
    supermassive black hole at the center.
  \end{abstract}
  


  \section{INTRODUCTION AND MOTIVATION}

  It is increasingly evident that the growth histories of different
  components of galaxies: stars, gas, and supermassive black holes,
  are intricately connected. Observations indicate a relationship
  between the masses of supermassive black holes (SMBHs) and various
  properties of their host galaxies, such as the spheroid mass
  \citep{Magorrian98} and the velocity dispersion of stars in the
  bulge \citep{FM00, Geb00, Trem02}. Recent simulations have shown
  that feedback from accreting SMBHs can regulate the growth of the
  black holes as well as the evolution of their host galaxies, making
  feedback a potentially important piece of the theory of SMBH-host
  galaxy co-evolution \citep[e.g.][]{DiMatteo05, Springel05,
  DiMatteo07, Sijacki07}.

  Building up the mass of a galaxy and driving AGN feedback requires a
  continuous replenishment of fuel in the center of the galaxy. A
  comprehensive understanding of such fueling is not possible without
  detailed knowledge about matter transport from large scales to the
  vicinity of the black hole. That transport helps determine, in
  particular, both the amount of material available for accretion,
  feedback, and star formation, as well as the strength and duration
  of the fueling. Large scale gravitational tidal fields created
  during major mergers of galaxies are thought to be effective at
  funneling matter toward the centers of galaxies. Indeed, simulations
  and semi-analytic modeling of hierarchical growth scenarios have
  shown that a combination of accretion and black hole mergers can
  effectively build-up the local black hole population
  \citep{KaufHae00,YooME04,VolRees05,Malbonetal06,VolRees06,Lietal07},
  as well as assemble the $\sim 10^9$ M$_{\sun}$ black holes already
  observed to be present in quasars at $z\approx6$
  \citep{Fan03,Lietal07}.

  Secular evolution of galaxies, particularly in the absence of major
  merger events, can drive fuel down to the center as well. Global bar
  instabilities are thought to be efficient at transporting angular
  momentum, allowing material in the disk to move toward the center of
  the galaxy
  \citep{Robertsetal79,Simkinetal80,Noguchi88,barswinbars,KormKenn04,RegTeu04}.
  In the ``bars within bars'' scenario of \citet{barswinbars}, a large
  scale galactic bar drives gas inward where it forms a
  self-gravitating disk. As this disk becomes unstable, a smaller
  secondary bar forms, driving material down to scales of order $10$
  pc, at which point other physics can take over and transport the
  material the rest of the way toward the black hole
  \citep[e.g.][]{ShlosRev90}.
  
  Given the complexity of transporting matter from cosmological scales
  all the way down to the circumnuclear region of a galaxy, and
  ultimately to the vicinity of a SMBH, it is clear that a thorough
  understanding of the relationship between SMBHs and galaxy formation
  requires modeling of a wide range of scales. Recently, cosmological
  SPH simulations have been combined with smaller scale simulations of
  a SMBH host galaxy environment to study the effects of merger driven
  fueling and AGN feedback on supermassive black hole growth and
  demography \citep{DiMatteo07, Sijacki07}. Reaching a large dynamic
  range in N-body+SPH simulations, \citet{Mayeretal07} have studied
  galaxy dynamics during mergers resulting in a supermassive black
  hole binary. Small scale simulations have addressed gas dynamics in
  sub-galactic scale disks with high resolution \citep{Fukuda00,
  Wada01, WadaNorman01,Escala07,WadaNorman07}, finding the
  development of a turbulent, multi-phase interstellar medium (ISM).

  The study we present here follows the evolution of the circumnuclear
  region in a typical galaxy environment rather than a dramatic, rare
  event such as a major merger, in a self-consistent cosmological
  simulation. We have specifically chosen a simulated galaxy that will
  evolve into a typical $L_*$ galaxy at $z=0$. The simulation follows
  the galaxy during a phase of its evolution in which it has not
  undergone any major mergers within several dynamical time scales, in
  order to follow the development of instabilities in the
  circumnuclear disk, which may drive the transport of matter and
  angular momentum from large to small scales. The cosmological
  simulations use the adaptive mesh refinement (AMR) technique to
  self-consistently model the gas dynamics in a single galaxy at high
  resolution (sub-pc resolution in the center of the galaxy). A large
  dynamic range ($> 10$ million), achievable with AMR technique,
  allows us to bridge cosmological scales to scales relevant for
  molecular cloud formation (the birthplace of stars) and AGN
  fueling. After studying the physics in this basic model galaxy, we
  can begin to include physical processes that are directly relevant
  to the problem of SMBH growth in the context of galaxy evolution,
  such as AGN feedback.

  It is a complex task to implement mergers, feedback and secular
  evolution in large, cosmological simulations all at once. Our
  approach is to split the problem into pieces to be addressed one at
  a time, ultimately building a more realistic simulation. While the
  present paper focuses on the structure of the galaxy, and the
  development of instabilities which may be responsible for driving
  matter and angular momentum transport within the galaxy, we plan to
  explore the time evolution of such transport, specifically the
  evolution of the accretion rates of mass and angular momentum, in a
  subsequent paper. The nature of the circumnuclear region and a
  quantitative analysis of accretion rates each have important
  consequences for galaxy evolution.

  The organization of the paper is as follows. In Section
  \ref{sec:sim}, we describe the details of the simulation and the
  adopted ``zoom-in'' method. In Section \ref{sec:res}, we describe
  the features of the highly-resolved galaxy in a single zoom-in
  episode at $z=4$, including a stability analysis of the disk and the
  potential role of physics not included in the current
  simulation. Section \ref{sec:miss} examines the potential role of
  physics missing from the simulations. Finally, Section
  \ref{sec:disc} contains a summary of our conclusions and their
  interpretation.


  \section{SIMULATION}
  \label{sec:sim}

  Sections \ref{subsection:art} and \ref{subsection:ref} briefly
  explain details of the simulation and of our methods. Additional
  details regarding tests of angular momentum conservation, and the
  treatment of dark matter particle discreteness effects will be
  available in the PhD thesis of the author Levine (also see the
  discussion of angular momentum conservation in Appendix
  \ref{sec:con}).
  
  \subsection{Simulation Code}
  \label{subsection:art}

  It is unfeasible to follow the evolution of a galaxy all the way
  down to the scale of a black hole hole accretion disk over
  cosmological times. Such a simulation spans a large range of scales
  (with a dynamic range $> 10^7$), and each spatial scale has a
  different temporal scale. As a result, the Courant condition for
  numerical stability requires very small time steps in the most
  highly refined regions, making it computationally expensive to
  follow the evolution of the galaxy over long periods of time. It is
  much more feasible to start with a lower resolution cosmological
  simulation, and zoom in to the small-scale region with increasingly
  smaller cell sizes and time steps. As the simulation evolves on
  small scales, the large scale portion of the simulation does not
  undergo much evolution, and does not need as high resolution.

  The cosmological simulations presented in this paper are conducted
  with the Adaptive Refinement Tree (ART) code \citep{Kravtsovetal97,
  KravtsovPhD, Kravtsovetal02}. The ART code includes the technique of
  adaptive mesh refinement (AMR), allowing high resolution of a galaxy
  residing in a small region of the cosmological simulation, while
  following the rest of the simulated region with lower
  resolution. The technique is appropriate for studying the structure
  and evolution of a galaxy over a large dynamical range and in a
  cosmological context.
 
  The ART code includes a range of physics for modeling dark matter,
  stars, and gas dynamics. Gas is converted into stars in cells with
  densities greater than $\rho_{\textrm{SF}}$ and temperatures less
  than $T_{\textrm{SF}}$, where $\rho_{\textrm{SF}} = 1.64
  \dim{M}_{\sun} \dim{pc}^{-3}$ and $T_{\textrm{SF}} = 9000$ K
  \citep[see][for more details]{Kravtsov03}, resulting in a star
  formation efficiency consistent with a Kennicutt law on kiloparsec
  scales \citep{Kennicutt98} and with observations on $100$ pc scales
  as well \citep[e.g.][]{Youngetal96, WongBlitz02}. The code follows
  ISM physics, such as molecular hydrogen formation, and gas cooling
  by heavy elements and dust under the assumption of collisional
  ionization equilibrium. The cooling and heating rates are tabulated
  as functions of gas density, temperature, metallicity, and redshift
  over the temperature range $10^2<T<10^9$ K using CLOUDY
  \citep{Cloudy98}, which accounts for the metallicity of the gas, and
  formation of molecular hydrogen and cosmic dust. In future studies,
  we plan to include the ART code's radiative transfer capabilities,
  which will be necessary for implementing AGN feedback.


  \subsection{``Zooming-In'' to the Center of a Galaxy}
  \label{subsection:ref}

  We begin with a cosmological simulation, evolved from a realization
  of a random Gaussian density field at $z=50$, with periodic boundary
  conditions, measuring $6 h^{-1}$ comoving Mpc across. The simulation
  was first run with low-resolution in order to select a galactic-mass
  halo for subsequent study. A Lagrangian region the size of five
  virial radii of the halo at $z=0$ was then identified at $z=50$, and
  re-sampled and run with higher resolution to $z=2.8$ \citep[see][for
  a description of the technique]{Klypinetal01}. The Lagrangian region
  of the simulation was automatically refined three levels, as a
  minimum. There are $2.64 \times 10^6$ dark matter particles in the
  Lagrangian region, each with mass $9.18 \times 10^5 h^{-1}
  \dim{M}_{\sun}$. Subsequent refinement and de-refinement in this
  region followed a dark matter mass criterion, in which a cell was
  refined if its total dark matter mass is greater than $1.8 \times
  10^6 h^{-1} \dim{M}_{\sun}$. At $z=4$, the highest matter density
  peaks in the simulation have a maximum resolution of $\approx 50$ pc
  (in physical units; corresponding to $9$ levels of refinement on top
  of a $64^3$ root grid). The largest halo has a total mass of $2
  \times 10^{11} \dim{M}_{\sun}$ at $z=4$, and it contains the
  progenitor of an $L^*$ spiral galaxy. This initial cosmological
  simulation was run including metal enrichment and energy feedback
  from supernova, and radiative transfer in addition to the physics
  described in Section \ref{subsection:art}.

  \begin{figure*}
    \center{\includegraphics[width=1.9\columnwidth]{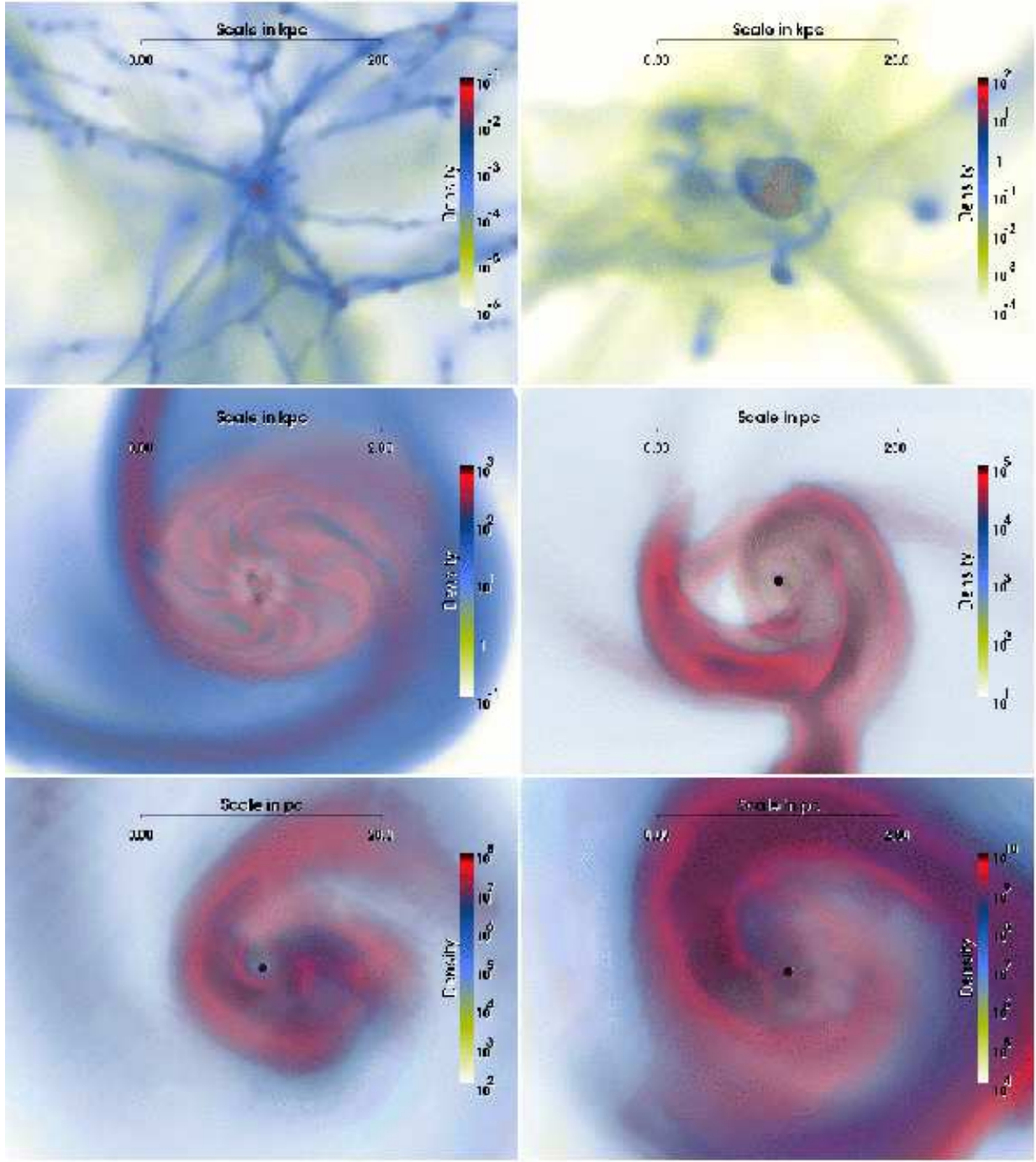}}
    \caption{\label{fig:pan} Volume rendering of the $3$-dimensional
    gas density on several different scales at $z=4$, from $200$ kpc
    in the upper left panel, to $2$ pc in the lower right panel. The
    color bar shows the density scale in units of
    $\dim{cm}^{-3}$. Note the presence of the spiral arms and
    instabilities on a wide range of scales. (This figure is best
    viewed in color.)} 
  \end{figure*}

  This particular $z=4$ cosmological simulation was chosen for this
  study because it contains a galaxy that has not been disturbed by a
  major merger since it merged with a galaxy with $25\%$ its mass at
  $z\approx6$. The time since this dynamically active period ($\approx
  600$ million years) is significantly longer than the dynamical time of
  the galactic disk at $z\approx4$ ($\approx 15$ million years). The
  galaxy is, however, far from quiescent, having grown in mass by
  $\approx 25\%$ from $z=5$ to $z=4$ via minor mergers and accretion
  from its environment.

  \begin{figure*}
    \center{\includegraphics[width=1.9\columnwidth]{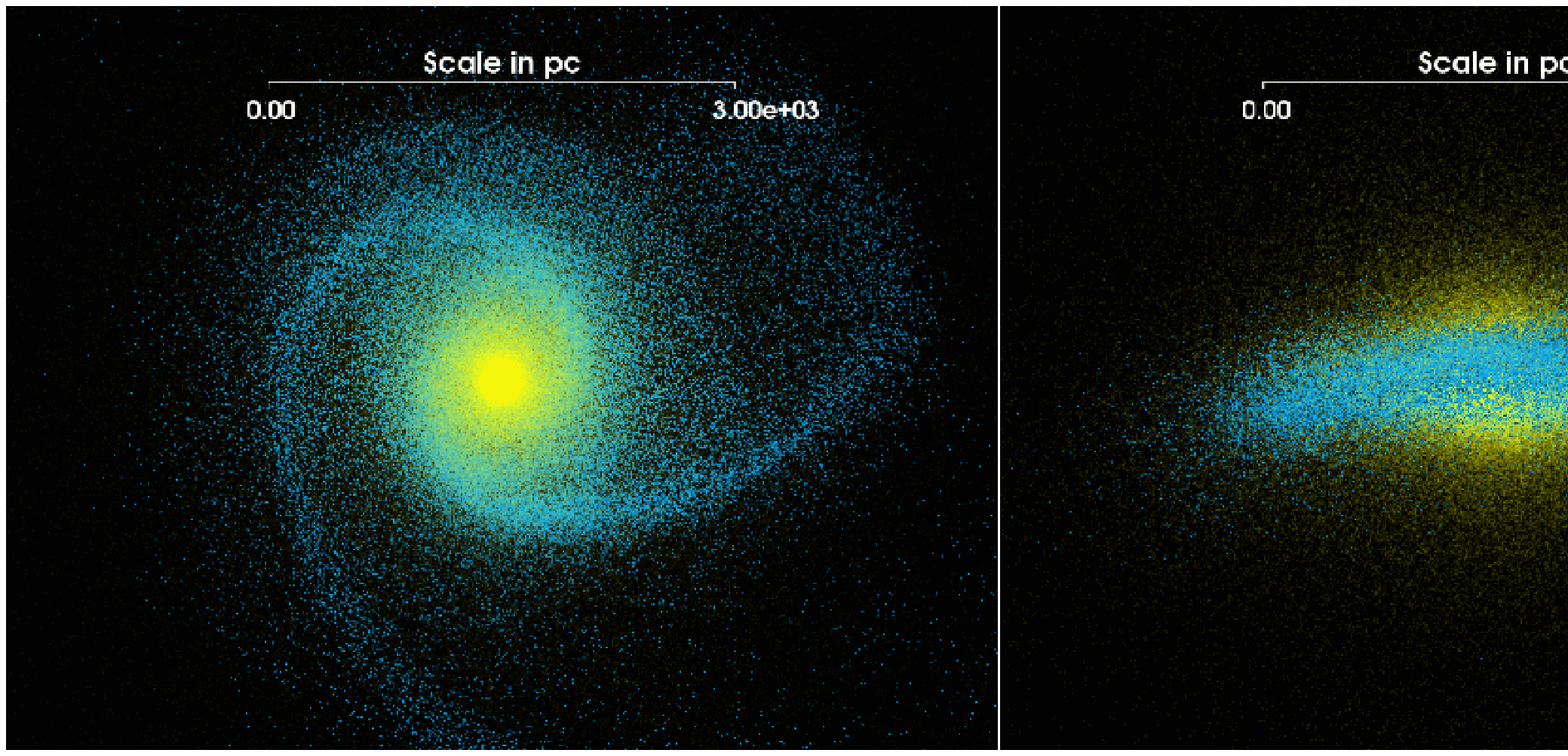}}
    \caption{\label{fig:star} The stellar component of the galaxy at
    $z=4$ on a scale of $3$ kpc. The stars in the image are divided
    into two populations, with yellow stars corresponding to the oldest
    stars in the galaxy, and blue stars corresponding to the youngest
    stars. The youngest stars are found primarily in the disk of the
    galaxy. (This figure is best viewed in color.)}
  \end{figure*}

  In the cosmological simulation, we allow further refinement (at $z
  \approx 4$), one level at a time, in a $1.5$ kpc region centered on
  the galaxy, effectively zooming in to the center of the galaxy with
  increasing resolution. The ``zoom-in'' technique is similar to the
  one used by \citet{Abeletal02} in simulations of the formation of
  the first star, which required an even larger dynamical
  range. Because we are starting simply in this first pilot study, and
  building a more realistic simulation in pieces, radiative transfer
  and stellar feedback have been switched off in the zoom-in
  portion of the cosmological simulation (they will be examined in
  future studies). The high-resolution portion of the simulation
  employs additional refinement criteria, refining according to a
  level-dependent mass criterion on levels $11$ and below\footnote{It
  should be noted that we adopt the convention of referring to level
  $0$ as the ``top'' level, and the maximum (most-refined) level as
  the ``bottom'' level, so that the terms ``above'' and ``below''
  refer to lower and higher resolution, respectively.} in the zoom-in
  region. The refinement criterion is defined so that the finer the
  resolution, the more aggressively the mesh refines, ensuring that
  there are enough highly refined cells to resolve structures on small
  scales. Specifically, the mesh refinement is super-Lagrangian in the
  circumnuclear region of the galaxy, and cells are marked for
  refinement if the gas mass in a cell is
  $m_{\textrm{r}}^{\textrm{level}-10}$ times the Lagrangian mass
  criterion ($9 \times 10^7 h^{-1} \dim{M}_{\sun}$), where
  $m_{\textrm{r}}=0.7$ is our fiducial value. The factor $0.7$ was
  chosen through experimentation, so as to make the refinement in the
  central region more aggressive, but not so aggressive that the
  simulation becomes too computationally expensive.

  During the zoom-in period, we increase the maximum level of
  refinement gradually, one level at a time, allowing the simulation
  to reach a quasi-stationary state on each level before moving to the
  next. The slow initial refinement allows the ART code to resolve
  spatial scales within the simulated galaxy while avoiding transient
  effects. Since the time steps depend on the sound speed of the gas,
  they are therefore related to the dynamical time, and a fixed
  minimum number of time steps on each level forces the mesh to evolve
  for several dynamical times. We have run parallel simulations with
  $100$, $300$, and $1000$ minimum steps on each level, and have
  determined that a requirement of a minimum of $300$ on each level
  sufficiently reduces transient numerical effects that result from
  refining too quickly.

  Throughout the initial zoom-in, we take precautions to avoid
  numerical artifacts.  As the resolution of the simulation increases
  and throughout its subsequent evolution, the density of dark matter
  is smoothed on scales corresponding to level $9$ resolution and
  above. This ensures that each cell contains $\approx 15-20$ dark
  matter particles, avoiding effects resulting from their
  discreteness.  It is well known that poor resolution in
  hydrodynamics codes can lead to artificial fragmentation in the
  gas. \citet{Truelove97} investigated these numerical instabilities
  and found that a simulation must resolve the Jeans length to avoid
  artificial fragmentation. In addition to the mass criteria described
  above, the ART code's refinement scheme meets the Jeans condition of
  \citet{Truelove97} on the maximum level of refinement (which
  increases gradually during the initial zoom-in episode), requiring
  that $\Delta x/\lambda_{\textrm{J}} < 0.25$, where $\Delta x$ is the
  resolution, or cell size, and $\lambda_{\textrm{J}}$ is the Jeans
  length. Additionally, as the mesh refines, rapidly collapsing
  structures can cause numerical instabilities at steep density
  gradients. The ART code implements artificial pressure support (as
  described in \citet{Machacek01}) on the maximum level of refinement,
  in order to avoid the over-collapsing of structures.

  After the initial zoom-in, the highest level cells are $0.03$ pc
  across (corresponding to $20$ levels of refinement), allowing us to
  determine the location for a supermassive black hole test particle
  with high precision, but without resolving the black hole accretion
  disk, which would require additional physics (such as
  magneto-hydrodynamics, or MHD) not currently included in the ART
  code. Our results extend reliably down to a scale of at least $0.12$
  pc, or the size of four level $20$ cells, which we consider to be
  our resolution limit. After reaching this maximum resolution (level
  $20$), we replace $3\times10^7 \dim{M}_{\sun}$ of the gas from the
  center of the simulated galaxy with a black hole point mass of equal
  mass and momentum. At present, we focus on the dynamical properties
  of the circumnuclear disk on scales where the gravity of the gas
  dominates that of the black hole. The addition of the black hole
  point mass here simply allows us to follow the black hole's location
  and velocity as a reference point. However the successful
  introduction of the black hole particle will be essential in
  subsequent simulations involving physics associated with the black
  hole (such as AGN feedback).

  After the initial zoom-in and the successful introduction of the
  black hole test particle, the simulation then evolves at high
  resolution for approximately one dynamical time at the $100
  \dim{pc}$ scale, showing a highly resolved galactic disk. For a
  quasi-Keplerian disk, the number of orbital periods,
  $N_{\textrm{orb}}$, undergone by the simulation at radii less than
  $100 \dim{pc}$, is $(R/100\dim{pc})^{-3/2}$. Figure \ref{fig:pan}
  shows the $3$-dimensional gas density in an approximately face-on
  view of the galactic disk, on several different spatial scales, for
  a single time step, demonstrating the large dynamic range of the
  simulation. The distribution of stars is shown in Figure
  \ref{fig:star}, in both a face-on and an edge-on view of the galaxy
  on a scale of $3$ kpc. The two different colors correspond to old
  and young stars (yellow and blue, respectively), with the youngest
  stars in the galaxy located primarily in the disk, and the oldest
  stars distributed more isotropically, populating the galaxy's bulge.


  \section{RESULTS}
  \label{sec:res}
  \subsection{Structure of the Circumnuclear Disk}
  \label{subsection:disk}

\begin{figure}[t]
    \centering \epsscale{1.2}  
    \plotone{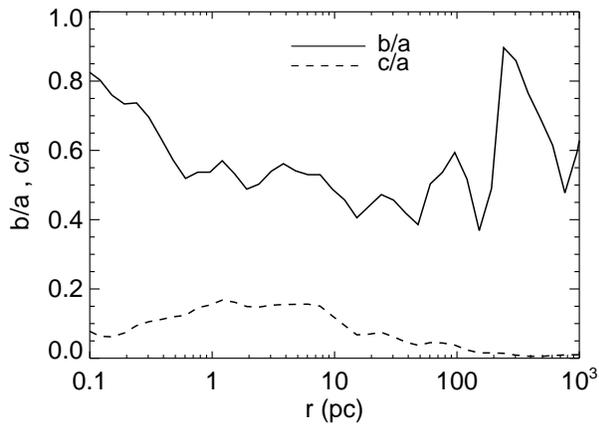}
    \caption{\label{fig:shape} Ratios of eigenvalues of the inertia
    tensor in spherical shells centered on the black hole. The ratio
    $c/a \ll b/a$, indicating a disk structure.}
  \end{figure}

  Volume rendered images of the gas density in the $z=4$ simulation,
  such as those in Figure \ref{fig:pan}, clearly show the spiral disk
  structure of the galaxy. The gas disk extends inward all the way to
  sub-pc scales, which are the smallest scales resolved in our
  simulation\footnote{In particular, the simulated disk does not
  contain the commonly observed toroidal structure in the inner few pc
  of the galaxy. The failure of our simulation to reproduce the AGN
  torus may be the result of additional physics still missing in the
  simulation, as we will discuss in \S \ref{subsection:torus}.}. The
  geometric structure of the circumnuclear region is best illustrated
  by the eigenvalues of the inertia tensor, calculated in spherical
  shells around the black hole. The inertia tensor is given by

  \begin{equation}\label{eq:iner}
    I_{jk} = \sum_{\alpha=1}^N m_{\alpha}x_j^{\alpha}x_k^{\alpha},
  \end{equation}

  \noindent where $m_{\alpha}$ is the gas mass of cell $\alpha$ at radius
  $r_{\alpha} = |\bf{x}_{\alpha}|$ inside a spherical shell containing
  $N$ cells, and centered on the black hole. The eigenvalues of the
  inertia tensor define the principal axes of each shell, and their
  ratios describe the shape. Figure \ref{fig:shape} shows the average
  radial profile of the ratios of the inertia tensor's
  eigenvalues. Above $0.1$ pc, $c$ is significantly smaller than $b$
  over several orders of magnitude in radius, indicating a disk
  structure. However, the fact that $b/a$ is significantly less than
  one in parts of the disk indicates that the disk is not axially
  symmetric, and that it has large-scale structures, such as spiral
  waves and bars.

\begin{figure}[t]
    \centering \epsscale{1} 
    \plotone{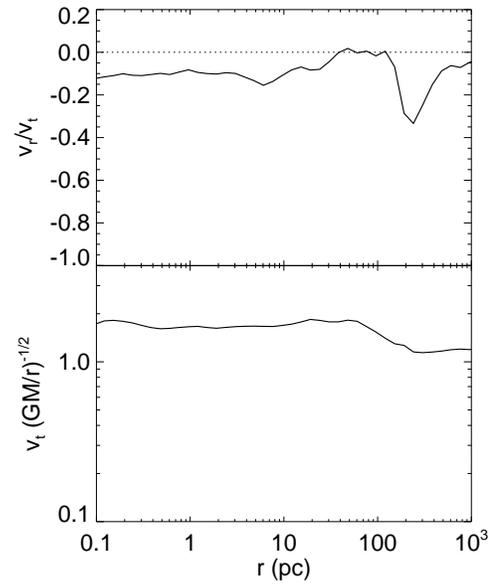}
    \caption{\label{fig:vel} {\it Top:} Ratio of radial and tangential
    velocity components of the gas, each averaged over $550{,}000$
    years. {\it Bottom:} Average tangential velocity component in units
    of the average quasi-Keplerian velocity.}
  \end{figure}

  The gas disk in the simulation is fully rotationally supported, with
  the tangential component of velocity dominating over the radial
  component by at least an order of magnitude, as is illustrated in
  Figure \ref{fig:vel}. The top panel of Figure \ref{fig:vel} shows
  the ratio of the radial to the tangential component of velocity,
  which is small throughout much of the disk, as the motion of the gas
  is almost entirely rotational. The average radial velocity is
  negative, indicating the inflow of gas. The bottom panel shows the
  tangential component of velocity in units of a quasi-Keplerian
  velocity, $\sqrt{GM(r)/r}$, determined by the interior total mass,
  $M(r)$, at each radius. The velocity is close to being Keplerian,
  but since the mass distribution in the disk resembles a flattened
  ellipsoid, rather than a spherical distribution, the rotation is
  slightly super-Keplerian.

\begin{figure}[t]
    \centering \epsscale{1.2} 
    \plotone{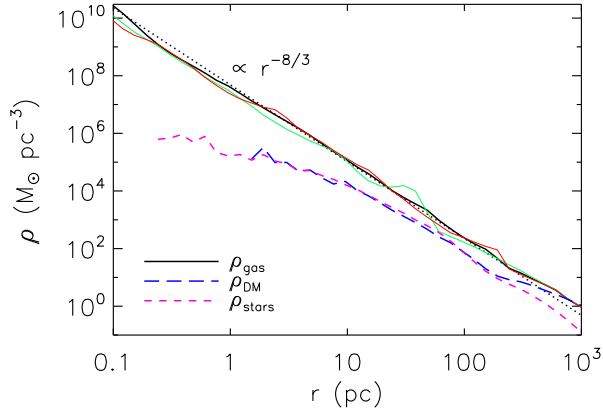}
    \caption{\label{fig:dens} Radial profiles of the gas, dark matter,
      and stellar mass density. Snapshots of the gas density are shown
      at $100{,}000$ and $500{,}000$ years after the initial
      refinement ({\it thin solid lines; green and red respectively,
      in the color version}), in addition to a $550{,}000$ year
      average ({\it thick solid line}) . The snapshots do not show
      much variation from the averaged profile, demonstrating that on
      timescales of several hundred thousand years, the disk is in a
      quasi-stationary state. The dotted line shows a power-law with
      slope $-8/3$ for comparison, which matches the gas density slope
      well. The dark matter and stellar mass density profiles are
      shown by the long-dashed ({\it blue in the color version}) and
      short-dashed ({\it magenta in the color version}) lines,
      respectively. In the inner $\sim 200$ pc, the slope of the dark
      matter and stellar density profiles is $\sim -2$ consistent with
      the adiabatic contraction model.}

  \end{figure}

  A remarkable feature of the circumnuclear disk is that the average
  gas density profile, measured within spherical shells centered on
  the black hole, follows an almost perfect power-law with little
  evolution in time. Figure \ref{fig:dens} shows the gas density
  profile from two different snapshots of the simulation, as well as
  the average of the profile over a $\approx 550{,}000$ year
  period. Both the snapshots and the average profiles of the gas
  density increase by $\approx 8$ orders of magnitude in the inner
  $100$ pc of the simulation, obeying a steep power-law with slope
  $-8/3$. The stability of the gas density profile indicates that the
  disk is in a quasi-stationary state on timescales of several hundred
  thousand years. Figure \ref{fig:dens} also shows the dark matter and
  stellar mass density profiles. In the central kiloparsec of the
  galaxy, gas comprises $\sim 62 \%$ of the mass ($\sim 2.3 \times
  10^{10} h^{-1} \dim{M}_{\sun}$ total in the central kpc), dominating
  the dynamics of the disk. In contrast, the stellar and dark matter
  populations comprise $\sim 13$ and $\sim 25 \%$ of the galaxy's mass
  inside the central kiloparsec. The disk is extremely gas rich at
  $z=4$, because the galaxy is still actively growing and has not yet
  formed all of its stars. Interestingly, we find that both the
  stellar and dark matter profiles measured in the central $\sim 200$
  pc of the galaxy follow simple $\propto r^{-2}$ power-laws. The
  profiles match those predicted for the adiabatic contraction of dark
  matter from an NFW profile \citep{NFW97}, using the model of
  \citet{Gnedinetal04}, all the way down to the $1\dim{pc}$ scale.

\begin{figure}[t]
    \centering
    \epsscale{1.2}
    \plotone{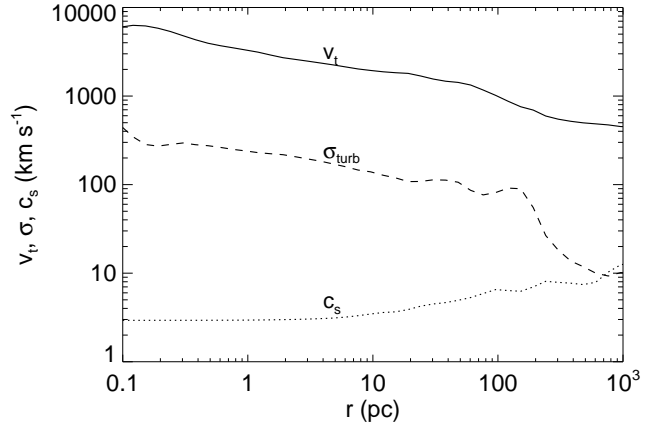}
    \caption{\label{fig:vcmp} A comparison of the radial profiles of
      the tangential velocity component of the gas ({\it solid}), its
      RMS velocity dispersion, or turbulent velocity ({\it dashed}),
      and the mean sound speed, computed in cells with $T<20{,}000$ K
      ({\it dotted}), time averaged over $550{,}000$ years. The sound
      speed is low, indicating cold, molecular gas within the central
      $20 \dim{pc}$.}
  \end{figure}

  Figure \ref{fig:vcmp} shows that the RMS velocity dispersion of the
  gas (measured between neighboring cells, and therefore depending on
  the resolution) greatly exceeds the sound speed in the inner $100$
  pc, indicating supersonic turbulence in the disk. Supersonic
  turbulence decays on a dynamical time scale, so the persistence of
  the turbulence in the circumnuclear disk of the galaxy indicates a
  driving mechanism, which will be addressed in the following
  sections. The mean sound speed of the gas indicates a cold,
  molecular gas disk within the central $100 \dim{pc}$ of the
  simulated galaxy. The floor in the sound speed shown in Figure
  \ref{fig:vcmp} is determined by the minimum temperature to which our
  adopted cooling rates (computed using the CLOUDY package) apply. The
  mean sound speed is computed for cells with $T<20{,}000$ K, because
  the spherical averages shown in Figure \ref{fig:vcmp} include the
  low density gas, infalling perpendicular to the plane of the
  galactic disk, and shock-heated to high temperatures ($\gtrsim 10^6$
  K).  Additionally, turbulence inside the disk produces localized
  shocks that briefly heat individual cells to similarly high
  temperatures. The inclusion of shock-heated gas produces a broad
  temperature distribution and raises the mean temperature of the gas
  so that it does not effectively describe the typical sound speed of
  the gas in the disk. Therefore, the mean is computed over a
  temperature range that selects cells in the disk, while excluding
  atypical cells, resulting in a more representative quantity for
  describing the thermal properties of the disk.

  \subsection{Angular Momentum Transport}
  \label{subsection:ang}

  The gas density profile of Figure \ref{fig:dens} motivates a simple
  analytic description of angular momentum transport in the
  disk. Averaged over a sufficiently long time, the disk can be
  considered to be approximately azimuthally symmetric. Additionally,
  the rotational velocity of the disk dominates over other velocity
  components: over the local velocity dispersion, and over the sound
  speed, as demonstrated in Figure \ref{fig:vcmp}. Therefore, the
  circumnuclear disk in the simulation is well approximated by a thin,
  rotationally supported, viscous, molecular gas disk. Conservation of
  angular momentum in such a disk is described by the following
  equation in cylindrical coordinates \citep{Pringle81}:

  \begin{equation}\label{eq:ang}
    \frac{\p}{\p t}(J_{\textrm{z}}) + \frac{1}{R}\frac{\p}{\p R}(R
    v_{\textrm{R}} J_z) = \frac{1}{2\pi R}\frac{\p G}{\p R},
  \end{equation}

  \noindent where the angular momentum density normal to the disk is
  given by $J_{\textrm{z}} = \Sigma R^2 \Omega$, where
  $\Omega=v_{\textrm{t}}/R$. The surface density of the gas is
  $\Sigma$, and $v_{\textrm{R}}$ is the radial component of velocity,
  which is small compared to the rotational velocity of the gas. The
  right hand side of equation (\ref{eq:ang}) describes the effects of
  a viscous torque, $G$, generated by turbulent motions of the
  gas. \citet{Pringle81} parameterizes the viscous torque $G$, which
  necessarily vanishes in solid body rotation, $\p \Omega / \p R = 0$,
  as

  \begin{equation}\label{eq:torqa}
    G(R,t) = 2 \pi \nu \Sigma R^3 \frac{\p \Omega}{\p R} 
  \end{equation}

  \noindent where $\nu$ is the coefficient of turbulent
  viscosity. During the initial zoom-in, the gas disk quickly reaches
  the power-law density profile shown in Figure \ref{fig:dens}. On
  short time scales of a few hundred thousand years, traced by a
  single high-resolution zoom-in episode of our simulation,
  time-averaged radial motions through the disk are small compared to
  the rotational motion, as shown in Figure \ref{fig:vel}. Over longer
  time scales, angular momentum is transported outward by a viscous
  torque driven by the turbulent motions of the gas, allowing
  accretion to slowly feed the black hole. The left hand side of
  equation (\ref{eq:ang}) can therefore be considered to be small on
  time scales of a few hundred thousand years. It then follows that
  the right hand side of equation (\ref{eq:ang}) should also be small,
  and that the torque $G$ is approximately constant with radius.

  Given that the gas density $\rho$ follows a power law (Figure
  \ref{fig:dens}), it is reasonable to approximate the surface density
  of the gas $\Sigma$ by a power law with the radius $R$:

  \begin{equation}\label{eq:surf}
    \Sigma \propto R^{-\beta}.
  \end{equation}

  \noindent Since the gravitational potential $\Phi$ satisfies
  $\nabla^2 \Phi = 4 \pi G \rho$ and $\Sigma = 2\rho R$, it follows
  that

  \begin{equation}\label{eq:phi}
    \Phi \propto R^{1-\beta}.
  \end{equation}

  \noindent Since the disk is almost entirely rotationally supported,
  the tangential velocity $v_{\textrm{t}}$ is effectively 
  the circular velocity $v_{\textrm{c}}$, defined by $v_{\textrm{c}}^2
  = R\ d\Phi/dR$. Thus, the angular speed is

  \begin{equation}
    \Omega = \frac{1}{R} \left(R \frac{d\Phi}{dR}\right)^{1/2} \propto
    R^{-(1+\beta)/2}, 
  \end{equation}

  \noindent and the viscous torque

  \begin{equation}\label{eq:torqb}
    G \propto R^{3(1-\beta)/2}\nu.
  \end{equation}

  \noindent If $\nu \propto R$ (which, as we will show in Section
  \ref{subsection:stab}, is the case in our simulations), then
  equation (\ref{eq:torqb}) implies a power-law slope of $\beta
  \approx 5/3$. It then follows that the volume density of the gas
  should scale as $\rho \propto R^{-8/3}$, consistent with the slope
  measured in Figure \ref{fig:dens}. Therefore, the simple description
  of angular momentum transport, maintained by turbulence in the disk,
  potentially provides a consistent interpretation of the power-law
  slope of the gas density profile of Figure \ref{fig:dens}.


  \subsection{Disk Stability and the Source of Turbulence}
  \label{subsection:stab}

\begin{figure*}
    \center{\includegraphics[width=1.9\columnwidth]{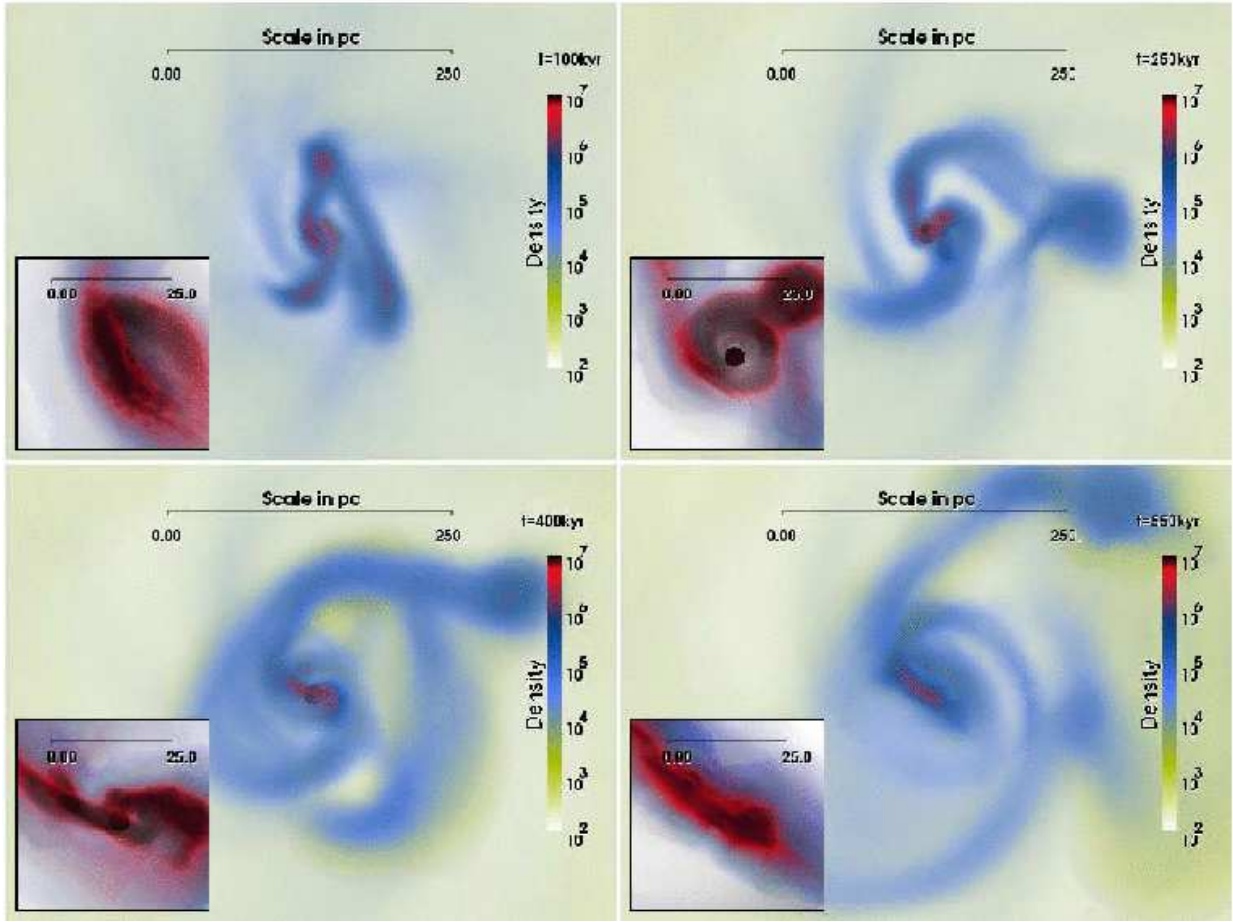}}
    \caption{\label{fig:evol} The evolution of global instabilities in
      the disk over time, on a scale of $\approx 125$ pc and on a scale
      of $\approx 12.5$ pc (inset). Each panel shows a volume rendering
      of the gas density (in $\dim{cm}^{-3})$ at a different
      epoch. The similarities between the appearance of the disk on
      the two different scales indicate the hierarchical structure of
      the disk. (This figure is best viewed in color.)} 
  \end{figure*}

  In Section \ref{subsection:disk}, we showed that the simulated
  galaxy contains a self-gravitating, turbulent, cold, molecular gas
  disk within the central $\approx 100 \dim{pc}$. Such disks are
  susceptible to instabilities and to fragmentation. In the snapshots
  shown in Figure \ref{fig:evol}, the presence of instabilities is
  illustrated by waves moving through the simulated disk. The
  snapshots trace the evolution of disk structure on scales of
  $\approx 125$ pc, and $\approx 12.5$ pc (inset). The rotational
  period of the disk at the outer radius is such that the disk has
  undergone at least one full rotation (several more at smaller radii)
  over the time scales followed by the present simulation. The panels
  show the somewhat chaotic formation and re-formation of spiral
  structures on these scales, in contrast with the more ordered spiral
  structure seen on kiloparsec scales (Figure \ref{fig:pan}). This
  section includes an analysis of the behavior of structures in the
  disk on the scales shown in Figure \ref{fig:evol}.

  The Toomre $Q$ parameter describes the stability of the disk against
  linear perturbations, and is given by $Q = \sigma_{\textrm{turb}}
  \kappa / \pi G\Sigma$, where $\kappa$ is the epicyclic frequency,
  and $\Sigma$ is the surface density of the gas, both determined
  locally \citep{Toomre64,GoldLB65}. Although the disk is cold, and
  the sound speed is low, the RMS velocity dispersion is substantially
  higher than the sound speed inside $100$ pc, indicating that the gas
  is turbulent. Therefore, in place of the sound speed typically used
  in the definition of the Toomre $Q$ parameter, we have substituted a
  turbulent velocity, given by the quadrature sum of the sound speed
  and the RMS velocity dispersion of the gas. The disk is, for the
  most part, quasi-Keplerian (as described in Section
  \ref{subsection:disk}), so that the epicyclic frequency $\kappa$ is
  proportional to the angular speed of the disk,
  $\Omega=v_{\textrm{t}}/r$.

\begin{figure}[t]
    \epsscale{1.2}
    \centering
    \plotone{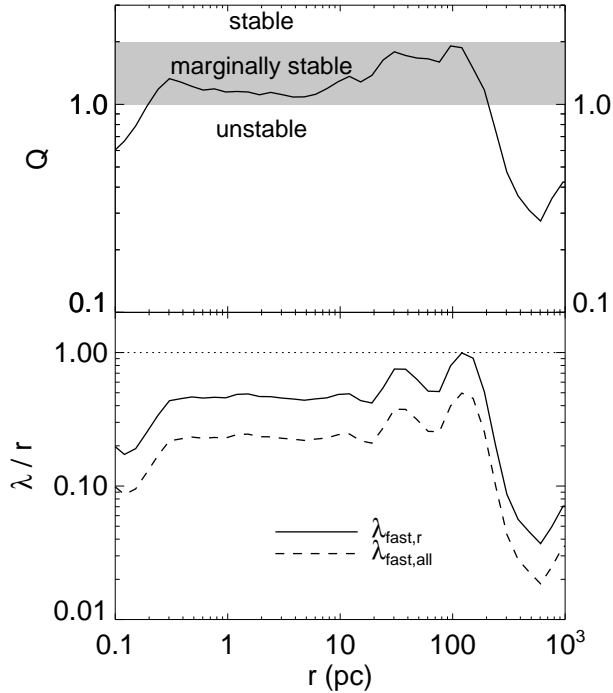}
    \caption{\label{fig:Q} {\it Top:} Toomre $Q$ parameter. For $Q<1$
      the disk is unstable, and for $Q\gtrsim2$ the disk is stable. In
      the regime $1<Q<2$, the disk is marginally stable (stable
      against axisymmetric modes, but not non-axisymmetric
      modes). {\it Bottom:} Average values of the the fastest growing
      unstable mode for axisymmetric perturbations,
      $\lambda_{\textrm{fast,r}}$ ({\it solid}), and the fastest
      growing unstable mode for all perturbations,
      $\lambda_{\textrm{fast,all}}$ ({\it dashed}), each divided by
      radius. The ratio $\lambda/r \gtrsim 0.1$ throughout the
      circumnuclear disk, so there is no catastrophic fragmentation
      down to small scales.}
  \end{figure}

  The top panel of Figure \ref{fig:Q} shows the Toomre $Q$ parameter
  for the simulated disk, corresponding to the region shown in Figure
  \ref{fig:evol}. Where $Q<1$, the disk is susceptible to local
  gravitational instabilities resulting from axisymmetric
  perturbations. For $Q>1$ the disk is likely to be stable against
  axisymmetric perturbations, but is still susceptible to
  instabilities arising from non-axisymmetric perturbations, which are
  less stable than radial perturbations \citep{Polyachenko}. For the
  density profile shown in Figure \ref{fig:dens}, the disk becomes
  stable for $Q\gtrsim2$. The shaded region in Figure \ref{fig:Q}
  ($1<Q<2$) therefore represents marginally stable values of the $Q$
  parameter, where the disk still might become unstable. Figure
  \ref{fig:Q} suggests that the disk in the simulated galaxy lies
  mostly in the region of marginal stability inside $\approx
  1\dim{kpc}$. In the case of axisymmetric perturbations, the fastest
  growing unstable mode corresponds to the scale
  $\lambda_{\textrm{fast,r}}$, at which $d\omega^2 / dk = 0$ (where
  $\omega$ is the angular frequency of waves in the disk), given by

  \begin{equation}\label{eq:fastr}
    \lambda_{\textrm{fast,r}} = \frac{2 \sigma_{\textrm{turb}}^2}{G
    \Sigma}.
  \end{equation}

  \noindent Using the condition for marginal stability from
  \citet{Polyachenko}, the fastest growing mode for all modes
  (axisymmetric and non-axisymmetric), $\lambda_{\textrm{fast,all}}$,
  is given by 

  \begin{equation}\label{eq:fasta}
    \lambda_{\textrm{fast,all}} = \frac{\lambda_{\textrm{fast,r}}}{2}
      = \frac{ \sigma_{\textrm{turb}}^2}{G \Sigma}.
  \end{equation}

  \noindent The bottom panel of Figure \ref{fig:Q} shows the ratio
  $\lambda / r$ for each of the fastest growing modes. In the inner
  $100$ pc, the scales of the fastest growing modes are smaller than
  the radius by only a factor of a few, implying that the disk is
  stable on scales $\lambda \ll R$ (at least in the linear
  regime). This is an indication that perturbations in the disk
  operate on a range of scales, but always on scales that are an
  appreciable fraction of the size of the system, driving global
  instabilities, which generate turbulence on smaller scales. The disk
  remains locally stable all the way into a region less than $1$ pc
  from the black hole. There is no catastrophic fragmentation into
  clumps small enough to form stars, so that angular momentum
  transport may continue uninterrupted by bursts of star formation.

\begin{figure}[t] 
    \centering
    \epsscale{1.2}
    \plotone{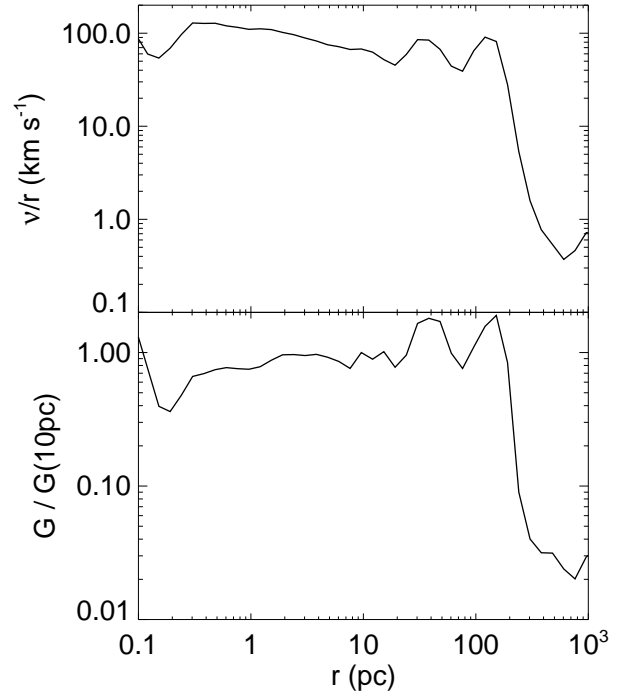}
    \caption{\label{fig:visc} {\it Top:} Ratio of kinematic viscosity
    $\nu$, given approximately by $\lambda_{\textrm{fast,r}}
    \sigma_{\textrm{turb}}$, to radius $R$. The ratio $\nu / R$ is
    roughly constant in radius. {\it Bottom:} The viscous torque $G$
    normalized to its $10$ pc value. The viscous torque is constant in
    radius, consistent with the power-law gas density slope $\rho
    \propto R^{-8/3}$.}
  \end{figure}

  By dimensional analysis, the turbulent kinematic viscosity $\nu$
  discussed in Section \ref{subsection:ang} can be described by
  $\lambda \sigma_{\textrm{turb}}$, where $\lambda$ is a
  characteristic scale for turbulence and $\sigma_{\textrm{turb}}$,
  the turbulent velocity shown in Figure \ref{fig:vcmp}, is a
  characteristic velocity. It is natural to identify the
  characteristic length scale with the length scale corresponding to
  the fastest growing unstable mode, $\lambda_{\textrm{fast,r}}$, so
  that $\nu \sim \lambda_{\textrm{fast,r}}
  \sigma_{\textrm{turb}}$. The top panel of Figure \ref{fig:visc}
  shows that the turbulent viscosity $\nu$, given approximately by
  $\lambda_{\textrm{fast,r}} \sigma_{\textrm{turb}}$, scales linearly
  with the radius over much of the disk, so that $\nu \propto
  R$. Although the behavior of the turbulent viscosity $\nu$ can
  potentially change with time, the viscous torque $G$ should remain
  approximately constant in radius for the above interpretation to be
  valid. The bottom panel of Figure \ref{fig:visc} shows the average
  viscous torque $G$, normalized to its value at $10$ pc. The
  normalized value is close to unity over most of the circumnuclear
  region, which is indeed the condition needed to explain the $-8/3$
  slope of the spherically-averaged gas density profile of Figure
  \ref{fig:dens}.

  \noindent Thus, a simple description of the structure of the
  circumnuclear disk based on equations (\ref{eq:ang}-\ref{eq:torqb})
  and Fig.\ \ref{fig:visc} provides a good match to the simulation
  results, supporting the idea that global instabilities in the disk
  are responsible for generating turbulence in the gas, resulting in
  the power-law slope described in Sections \ref{subsection:disk} and
  \ref{subsection:ang}.

  The conditions in the circumnuclear disk are potentially conducive
  to gaseous bar formation, the conditions for which have been studied
  extensively for different geometries and physical conditions,
  resulting in a variety of criteria
  \citep[e.g.][]{OP73,Efstat82,Christo95_1,Christo95_2,Bottema03,Wyse04}.
  A comparison with the criterion of \citet{OP73}, which characterizes
  stability in terms of the ratio $t=T/|W|$ of kinetic to
  gravitational potential energy, indicates that the disk is both
  secularly and dynamically unstable to bar formation on all scales $r
  \gtrsim 0.1 \dim{pc}$ adequately resolved by the simulation. Because
  of the chaotic and transient behavior of the instabilities, the
  structures shown in Figure \ref{fig:evol} do not resemble a single,
  well defined bar, but rather a highly perturbed ``bar.''

\begin{figure}[t] 
    \centering
    \epsscale{1.2}
    \plotone{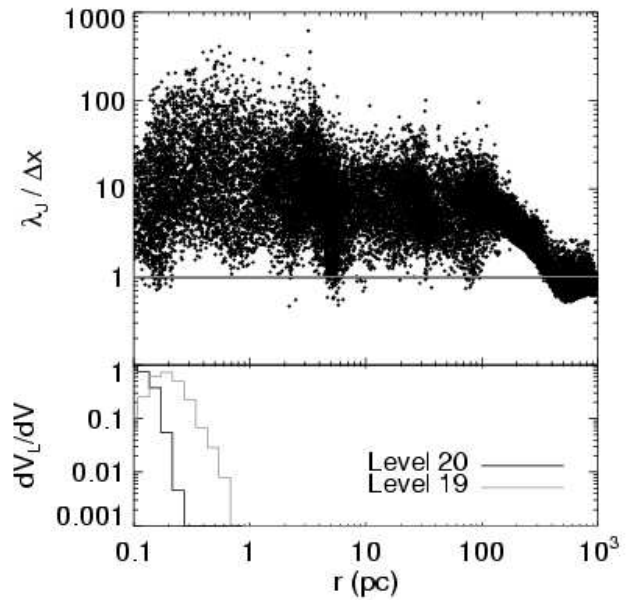}
    \caption{\label{fig:jeans} {\it Top:} Scatter plot showing the
      ratio of the local Jeans length, $\lambda_{\textrm{J}}$, to the
      cell size, $\Delta x$ at $300{,}000$ years after the initial
      refinement. Only cool cells with temperatures less than $10^3$ K
      are shown. {\it Bottom:} Histogram showing the volume of level
      $20$ and of level $19$ cells as a function of radius. The
      simulation only refines to the maximum levels in the center of
      the circumnuclear disk.}
  \end{figure}

  Since the Toomre criterion applies only to small linear
  perturbations and the Ostriker-Peebles criterion describes global
  modes, the question remains whether non-linear effects lead to
  fragmentation on smaller scales, below the resolution of the
  simulation. The top panel of Figure \ref{fig:jeans} shows the ratio
  of the local Jeans length of the gas, $\lambda_{\textrm{J}}$, to
  cell size, $\Delta x$, for all simulation cells in the circumnuclear
  disk (with temperatures less than $10^3$ K). In most cells, the
  Jeans length is larger than the cell size, preventing the gas from
  fragmenting into sub-cell sized clouds, ultimately leading to star
  formation. The Truelove criterion for preventing numerical
  fragmentation is enforced on the maximum level of refinement (level
  $20$) only, so there are no numerical restrictions to prevent gas
  from collapsing all the way to level $20$. However, only the central
  sub-pc part of the disk reaches levels $19$ and $20$, as shown by
  the histogram in the lower panel of Figure \ref{fig:jeans},
  indicating that the disk is stable to non-linear effects. While some
  cells in the disk have $\lambda_{\textrm{J}} < \Delta x$ and may
  form stars, Figure \ref{fig:jeans} demonstrates that most of the
  disk is stable against collapse, and that there is no widespread
  burst of star formation. The few cells with
  $\lambda_{\textrm{J}}/\Delta x < 1$ may correspond to resonances
  where the pattern speed of waves in the disk is the same as the
  rotational velocity, leading to higher gas densities, and possible
  star formation. However, a detailed analysis of the behavior of
  these resonances with time is beyond the scope of this paper.

  \subsection{The Nature of the Turbulence}
  \label{subsection:turb}

  Turbulence has been widely studied in modeling of the ISMs of
  galaxies and star forming regions \citep[for recent reviews,
  see][]{MacLowKlessen04,McKeeOst07}. It is important to understand
  the nature of the turbulence in the present simulated galaxy as it
  plays a key role in the disk properties. The global instabilities
  arising in the simulated circumnuclear disk generate turbulence on a
  range of scales, maintaining the quasi-steady state of the disk.

\begin{figure}[t] 
    \centering
    \epsscale{1.}
    \plotone{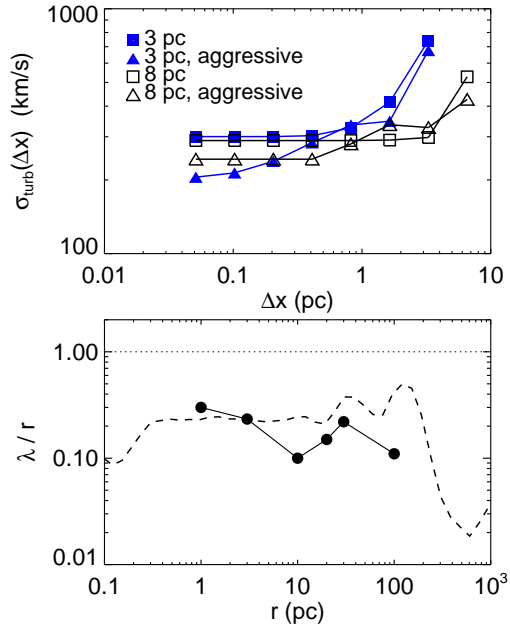}
    \caption{\label{fig:spec} {\it Top:} Turbulent velocity as a
    function of scale (or cell size) at $3$ and $8$ pc, for the
    fiducial and aggressive refinement runs. The more aggressive
    refinement run resolves the scaling of the turbulent down to
    smaller scales. {\it Bottom:} A comparison of the smallest
    resolved scale of turbulence and the fastest growing unstable
    mode, $\lambda_{\textrm{fast,all}}$ (where
    $\lambda_{\textrm{fast,all}}$ is the same as in Figure
    \ref{fig:Q}).}
  \end{figure}

  Turbulence dissipates energy in such a way that the turbulent
  velocity scales as $\sigma_{\textrm{turb}} \propto \lambda^q$, where
  $q=1/3$ for incompressible, sub-sonic (Kolmogorov) turbulence and
  $q=1/2$ for compressible turbulence in the zero-pressure limit
  (Burgers turbulence). The supersonic turbulence in the simulated
  disk falls between these two limits. In AMR simulations, it is
  straightforward to measure the turbulent velocity on different
  scales because the information on different levels of refinement is
  readily available. However, the measurements are only accurate to
  within a factor of two in scale because the cell size decreases by a
  factor of two with each refinement. The approximate scaling of the
  turbulence is shown in the top panel of Figure \ref{fig:spec}. The
  figure shows the turbulent velocity (given by the RMS velocity
  dispersion between neighboring cells) at two different radii for two
  different simulation runs. The first is the fiducial run described
  in the previous sections, and the second is a short portion of the
  fiducial run, with a more aggressive refinement criterion on levels
  $13$ and below, given by $m_{\textrm{r}}^{\textrm{level}-10}
  \textrm{max}[0.5^{\textrm{level}-12},\ 0.125]$ (where
  $m_{\textrm{r}}$ is an empirically determined parameter defined in
  Section \ref{subsection:ref}). The more aggressive run probes
  smaller scales at a given radius, in order to capture the scale of
  the turbulence. The turbulent velocities shown in Figure
  \ref{fig:spec} demonstrate the scaling of the turbulence down to
  small scales. It is difficult to determine the precise slope of the
  scaling using measurements from the present simulation because of
  the limitation imposed by the refinement scheme. While a more
  precise characterization of the turbulence spectrum is beyond the
  scope of the present simulation, Figure \ref{fig:spec} shows that
  the slope approximately falls between the Kolmogorov and Burgers
  turbulence limits, as expected.

  At scales below the resolution at a given location in the disk, the
  turbulent velocities in Figure \ref{fig:spec} level off at constant
  values because the turbulence scaling can only be measured down to
  the resolution limit. The more aggressive refinement run ({\it
  triangles} in Figure \ref{fig:spec}) probes smaller scales for a
  given radius, and therefore levels off at smaller scales. The bottom
  panel of Figure \ref{fig:spec} shows the approximate scale at which
  the flattening of the spectrum occurs for several different radii in
  the circumnuclear disk of the fiducial run. For comparison, the
  fastest growing unstable mode, $\lambda_{\textrm{fast,all}}$ is
  shown as well. The turbulent velocity measured at the resolution of
  the simulation is a numerical quantity, but it roughly corresponds
  to the turbulence on the scale of $\lambda_{\textrm{fast,r}}$, which
  is a physical quantity. Figure \ref{fig:spec} demonstrates that the
  simulation resolves all scales that are Toomre $Q$ unstable,
  supporting the resulting interpretation that instability driven
  turbulence maintains the quasi-stationary state of the circumnuclear
  disk.

\begin{figure}[t]
    \centering
    \epsscale{1.2}
    \plotone{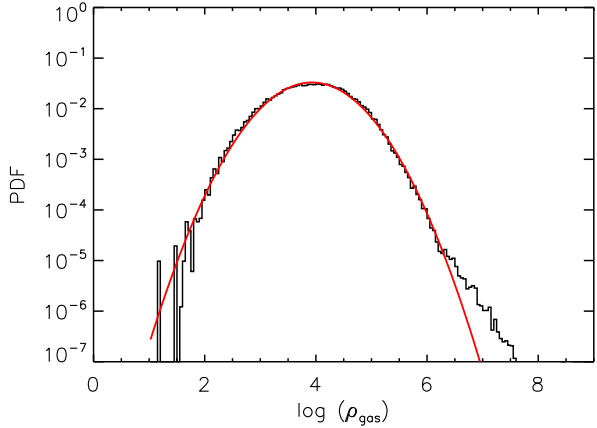}
    \caption{\label{fig:pdf} A volume-weighted PDF of the gas density
      in cells at a scale of $\sim 30$ pc, with temperatures $< 10^3$
      K (in order to exclude the hot corona outside the disk),
      averaged over $\sim 550,000$ years. The PDF is well fit by a
      log-normal distribution over the range $10^2 \lesssim \rho
      \lesssim 10^6 \dim{M}_{\sun} \dim{pc}^{-3}$ with a mean density
      of $10^{3.9} \dim{M}_{\sun} \dim{pc}^{-3}$ ({\it line}).}
  \end{figure}

  The density structure of the disk is also consistent with the
  description of turbulence in the disk. Supersonic turbulence
  typically imposes a log-normal density distribution on the gas, as
  demonstrated by models of turbulence in molecular clouds
  \citep[e.g.][]{Vazquez94, PassotVaz98, Wada01, Kritsuketal07,
  McKeeOst07}, such as

  \begin{equation}\label{eq:pdf}
    \frac{dV}{d\rho} = \frac{1}{\rho \sqrt{2\pi\sigma^2}} \
    \textrm{exp}\left[\frac{-(\textrm{ln} \rho - \mu)^2}{2
    \sigma^2}\right],
  \end{equation}

  \noindent where the mean, $\mu$, is defined as
  $\overline{\textrm{ln}\ \rho}$, and $\sigma$ is the
  dispersion. Figure \ref{fig:pdf} shows the volume-weighted
  probability distribution function (PDF), time-averaged over $\sim
  250,000$ years, for a shell of gas at a radius of $30$ pc. The PDF is
  well fit by a log-normal distribution over at least $4$ orders of
  magnitude in density, consistent with models of supersonic turbulence.

  \section{Possible Effects of Missing Physics}
  \label{sec:miss}

  The simulations presented here do not include all of the potentially
  important physics for galaxy evolution. Therefore, in this section
  we discuss the potential effect that the physics missing from the
  simulations might have on our results.

  \subsection{Optically Thick Cooling}
  \label{subsection:torus}

\begin{figure}[t] 
    \centering \epsscale{1.2} \plotone{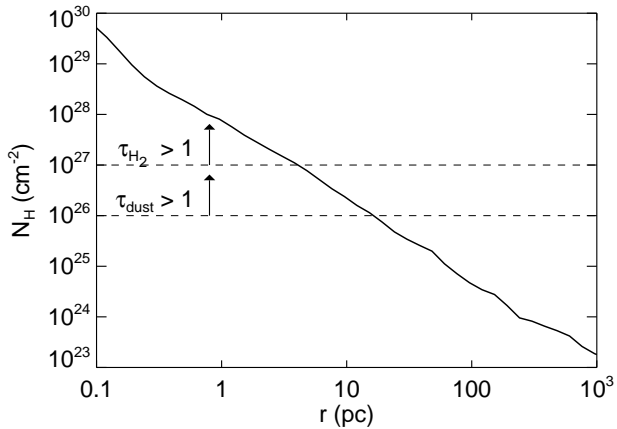}
    \caption{\label{fig:col} Column density of the gas. The horizontal
    lines show the column densities at which dust and H$_2$ become
    optically thick to their own cooling radiation.}
  \end{figure}

  Cooling in the high density, central region of the galaxy is not
  treated entirely correctly by the ART code. The column density of
  the molecular gas in the center is so high that the gas is expected
  to become optically thick to its own cooling radiation
  \citep[e.g.][]{RipAbel04}. Additionally, the presence of dust grains
  in this region traps radiation and halts cooling further
  \citep[see][]{DraineLee84, Ossenkopf94}. Figure \ref{fig:col} shows
  the column density vertically through the circumnuclear gas
  disk. Horizontal lines are shown for comparison, corresponding to
  the column densities at which dust and H$_2$ each become optically
  thick to their own cooling radiation ($\sim 10^{26} \dim{cm}^{-2}$
  and $\sim 10^{27} \dim{cm}^{-2}$, respectively, at $0.1$ solar
  metallicity, which is the metallicity of the circumnuclear disk in
  the simulation at $z=4$). In the inner $\approx 10$ pc, the column
  density of the gas is large enough that the opacity of dust and
  molecular gas must be accounted for to accurately describe cooling
  in the simulation.

  The simulation runs presented here do not include optically thick
  radiative transfer. This may explain why the gas remains in a thin
  disk all the way in to sub-pc scales, and does not resemble the
  obscuring tori observed in the inner few parsecs of AGN. Should the
  opacity of the gas to its own cooling radiation be included, the gas
  around the mid-plane of the disk would not be able to cool and would
  heat to temperatures corresponding to the energy dumped into the gas
  by the turbulence. These higher temperatures may be able to provide
  a substantial vertical pressure support in the central few parsecs
  of the disk, resulting in a thicker, more toroidal-like
  structure\footnote{Notice, that the outer, optically thin layers of
  such a disk would continue cooling efficiently, covering the hot
  interior with a cold molecular ``skin''. Such a skin may become
  Rayleigh-Taylor unstable, leading to fragmentation of the torus into
  individual clouds.}. However, it has been suggested that
  hydromagnetic disk winds, and not hydrostatic pressure support, may
  be responsible for sustaining the optically and geometrically thick
  obscuration region, or ``obscuring torus,'' in the nuclei of
  galaxies \citep[e.g.][]{KonigKartje94,ElitzShlos06}. In which case,
  the inclusion of optically thick radiative transfer may not be
  sufficient to create a ``torus'' in our simulations.

  In future simulations we plan to remedy this limitation of the
  present simulation by accounting for the opacity of the high 
  density region in the center, which will allow us to consistently
  incorporate radiative feedback from a central source and to test the
  above hypothesis.

  \subsection{Magnetic Fields}
  \label{subsection:mag}

  The ART code does not include magnetic fields, whereas MHD is
  certainly important for accretion disk physics \citep[see][and
  references therein]{BalbusHawley98}. It is for this reason that the
  present study has been restricted to scales larger than $\approx 0.1
  \dim{pc}$, corresponding to about $10^4$ Schwarzschild radii for a
  black hole of mass $3 \times 10^7$ M$_{\sun}$.

\begin{figure}[t] 
    \centering
    \epsscale{1.2}
    \plotone{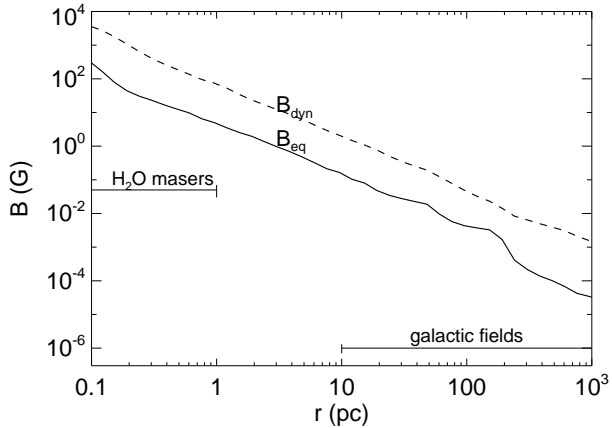}
    \caption{\label{fig:B} Equipartition ({\it solid}) and dynamical
    ({\it dashed}) magnetic fields needed to influence the gas
    dynamics in the disk.}
  \end{figure}

  At scales larger than the accretion disk, the absence of magnetic
  fields in the ART code is probably not important. An estimate of the
  strength that an equipartition magnetic field would need in order to
  affect the disk on small scales is given by $B_{\textrm{eq}}^2
  \approx 4\pi \rho \sigma_{\textrm{turb}}^2$. In order to affect the
  large-scale dynamics of the gas, the magnetic field would have to
  have an even larger strength of order $B_{\textrm{dyn}}^2 \approx
  4\pi \rho v_{\textrm{t}}^2$. Figure \ref{fig:B} shows estimates of
  $B_{\textrm{eq}}$ and $B_{\textrm{dyn}}$, in order to demonstrate
  how high the magnetic field would need to be to significantly
  influence gas dynamics in the simulated galaxy. For most of the
  galactic disk, the above estimates for the magnetic field are far
  higher than the few $\mu$G fields observed in real galaxies
  \citep[e.g.][]{ZweibHeil97,Beck01}. Even in the sub-pc region, where
  water maser observations indicate stronger fields of a few tens of
  mG \citep[e.g.][]{Modjazetal05,Vlemmetal07}, the magnetic field is
  still too low to affect the gas dynamics in the model galaxy. We
  therefore conclude that magnetic fields, which are not included in
  our simulations, will have a negligible effect on the dynamical
  state of this disk unless their strength greatly exceeds
  observational measurements.

  \section{DISCUSSION AND CONCLUSIONS}
  \label{sec:disc}

  We have used a large-dynamic range cosmological simulation to study
  gas dynamics in the circumnuclear disk of typical mass galaxy at
  $z\approx4$ (evolving into an $L_*$ galaxy at $z=0$), resolving the
  distribution of matter from megaparsec scales all the way down to
  sub-pc scales (with $20$ levels of refinement). The simulation
  reveals a cold, fully molecular, self-gravitating, and turbulent
  rotationally-supported gas disk, which is globally unstable but
  locally stable.

  The global instability, operating on a range of scales comparable to
  the size of the system, generates turbulence down to the smallest
  scale resolved in the simulation. On small scales, the turbulence
  supports most of the disk against gravitational fragmentation and
  collapse. The disk, therefore, remains locally stable and reaches a
  quasi-stationary state. In that state, global instability drives
  bar-like and spiral-wave-like structures on time scales of the order
  of $100{,}000$ years at $100\dim{pc}$ scales (and on proportionally
  shorter time scales at smaller radii), but on a $500{,}000$ year
  time-scale the structure of the disk remains quasi-steady.

  The disk develops a power-law gas density profile with a well
  defined slope of $-5/3$ in surface density and $-8/3$ in
  spherically-averaged volume density. This slope is a natural
  consequence of the dynamical state of the disk: the turbulence,
  generated by the global instability, redistributes the angular
  momentum in the disk on a dynamical time scale, reducing the
  gradient of the viscous torque. The resulting (quasi-)steady state
  uniquely determines the gas density profile we find in the
  simulation and the distribution of the angular momentum with gas
  mass.

  The turbulence in the disk drives the local outward transport of
  angular momentum and the inward flow of gas toward the supermassive
  black hole on time scales of at least 10 million years, which are
  too long to follow in a single zoom-in episode of the
  simulation. Thus, we only capture a single snapshot in the
  cosmological life of the supermassive black hole. In follow-up
  work, we plan to consider several snapshots taken at different
  cosmological times in order to describe the system on longer time
  scales.

  The dynamical state of the disk that we find in our simulation
  appears to be consistent with the results of previous simulations of
  isolated circumnuclear disks in galaxies \citep{Fukuda00,
  Wada01, WadaNorman01, Escala07}. A distinctive feature of our
  approach is that we follow the dynamics of the circumnuclear disk
  within cosmological simulations. While most of the volume of the
  simulation evolves little on time scales relevant for the dynamics
  of the circumnuclear disk, cosmological scales provide realistic
  boundary conditions for the dynamics on sub-kpc scales. Since we
  find that the circumnuclear disk rapidly reaches a quasi-stationary
  state, its evolution is entirely governed by the boundary
  conditions. For the same reason, the fact that the cosmological
  simulation that was used for the initial conditions did not resolve
  the scale of the circumnuclear disk, does not compromise our
  results: the quasi-stationary state of the disk does not depend on
  the initial conditions, and so the lack of power on scales below
  about $100\dim{pc}$ in the initial cosmological simulation is not
  important.

  The adopted approach of this work is comparable to the recent study
  by \citet{Mayeretal07}, who used a cosmological simulation to model the
  coalescence of two supermassive black holes. While the detailed
  treatment of gas physics and the scientific questions answered by
  the two studies are different, many of our results are consistent
  with those of \citet{Mayeretal07}. For example, they find a similar
  slope for the spherically averaged density profile, although they do
  not elaborate on the physical mechanisms for the formation of this
  profile. 

  The local stability of the disk, supported by highly supersonic
  turbulence, implies that the disk is capable of continuously feeding
  a central black hole, uninterrupted by catastrophic bursts of star
  formation, which could have consumed the available fuel were the
  disk locally unstable. This interesting dynamical state of the disk
  provides a potential solution to the problem of how AGN fueling is
  maintained by self-gravitating gas disks \citep[for a related
  discussion, see e.g.][]{ShlosBeg89,Riceetal05,NayakshinKing07}.

  The circumnuclear disk in the simulation extends all the way to the
  maximally resolved scale of $\approx0.1\dim{pc}$, which corresponds
  to the outer part of the black hole accretion disk. We find no
  toroidal-like structures on several parsec scales, which are
  commonly inferred to exist around AGN. A possible reason that the
  disk does not form an AGN torus on the appropriate scales (if it
  indeed, should) is the limitation imposed by our implementation of
  gas cooling in the code. The ART code, like all existing
  cosmological codes, assumes that the cooling radiation from cosmic
  gas escapes freely into the IGM. This assumption breaks down at the
  densities and temperatures reached by the simulation in the inner
  $10\dim{pc}$. At this scale the disk becomes optically thick to its
  own cooling radiation from dust and molecular hydrogen. In that
  regime the disk may to heat up and acquire a substantial
  amount of pressure support, which may result in a puffier, more
  toroidal-like configuration for the inner several parsecs of the
  disk. A consistent treatment of a putative AGN torus will require
  simulations that incorporate optically-thick cooling.

  \acknowledgements

  The authors thank Oleg Gnedin for his analysis of the mass profiles
  from the simulations. The authors also thank Mitch Begelman, Moshe
  Elitzur, Lucio Mayer, Brant Robertson, Isaac Shlosman, Volker
  Springel, Marta Volonteri, \& Keiichi Wada for comments on the
  paper. This work was supported in part by the DOE and the NASA grant
  NAG 5-10842 at Fermilab and by the NSF grants AST-0134373 and
  AST-0507596. Supercomputer simulations were run on the IBM P690
  array at the National Center for Supercomputing Applications and San
  Diego Supercomputing Center (under grant AST-020018N).

  
  \bibliography{ms}

\begin{thebibliography}{70}
\expandafter\ifx\csname natexlab\endcsname\relax\def\natexlab#1{#1}\fi

\bibitem[{{Abel} {et~al.}(2002){Abel}, {Bryan}, \& {Norman}}]{Abeletal02}
{Abel}, T., {Bryan}, G.~L., \& {Norman}, M.~L. 2002, Science, 295, 93

\bibitem[{{Balbus} \& {Hawley}(1998)}]{BalbusHawley98}
{Balbus}, S.~A., \& {Hawley}, J.~F. 1998, Reviews of Modern Physics, 70, 1

\bibitem[{{Beck}(2001)}]{Beck01}
{Beck}, R. 2001, Space Science Reviews, 99, 243

\bibitem[{{Bottema}(2003)}]{Bottema03}
{Bottema}, R. 2003, \mnras, 344, 358

\bibitem[{{Christodoulou} {et~al.}(1995{\natexlab{a}}){Christodoulou},
  {Shlosman}, \& {Tohline}}]{Christo95_1}
{Christodoulou}, D.~M., {Shlosman}, I., \& {Tohline}, J.~E. 1995{\natexlab{a}},
  \apj, 443, 551

\bibitem[{{Christodoulou} {et~al.}(1995{\natexlab{b}}){Christodoulou},
  {Shlosman}, \& {Tohline}}]{Christo95_2}
---. 1995{\natexlab{b}}, \apj, 443, 563

\bibitem[{{Di Matteo} {et~al.}(2007){Di Matteo}, {Colberg}, {Springel},
  {Hernquist}, \& {Sijacki}}]{DiMatteo07}
{Di Matteo}, T., {Colberg}, J., {Springel}, V., {Hernquist}, L., \& {Sijacki},
  D. 2007, ArXiv e-prints, 705

\bibitem[{{Di Matteo} {et~al.}(2005){Di Matteo}, {Springel}, \&
  {Hernquist}}]{DiMatteo05}
{Di Matteo}, T., {Springel}, V., \& {Hernquist}, L. 2005, \nat, 433, 604

\bibitem[{{Draine} \& {Lee}(1984)}]{DraineLee84}
{Draine}, B.~T., \& {Lee}, H.~M. 1984, \apj, 285, 89

\bibitem[{{Efstathiou} {et~al.}(1982){Efstathiou}, {Lake}, \&
  {Negroponte}}]{Efstat82}
{Efstathiou}, G., {Lake}, G., \& {Negroponte}, J. 1982, \mnras, 199, 1069

\bibitem[{{Elitzur} \& {Shlosman}(2006)}]{ElitzShlos06}
{Elitzur}, M., \& {Shlosman}, I. 2006, \apjl, 648, L101

\bibitem[{{Escala}(2007)}]{Escala07}
{Escala}, A. 2007, ArXiv e-prints, 705

\bibitem[{{Fan} {et~al.}(2003){Fan}, {Strauss}, {Schneider}, {Becker}, {White},
  {Haiman}, {Gregg}, {Pentericci}, {Grebel}, {Narayanan}, {Loh}, {Richards},
  {Gunn}, {Lupton}, {Knapp}, {Ivezi{\'c}}, {Brandt}, {Collinge}, {Hao},
  {Harbeck}, {Prada}, {Schaye}, {Strateva}, {Zakamska}, {Anderson},
  {Brinkmann}, {Bahcall}, {Lamb}, {Okamura}, {Szalay}, \& {York}}]{Fan03}
{Fan}, X., {Strauss}, M.~A., {Schneider}, D.~P., {Becker}, R.~H., {White},
  R.~L., {Haiman}, Z., {Gregg}, M., {Pentericci}, L., {Grebel}, E.~K.,
  {Narayanan}, V.~K., {Loh}, Y.-S., {Richards}, G.~T., {Gunn}, J.~E., {Lupton},
  R.~H., {Knapp}, G.~R., {Ivezi{\'c}}, {\v Z}., {Brandt}, W.~N., {Collinge},
  M., {Hao}, L., {Harbeck}, D., {Prada}, F., {Schaye}, J., {Strateva}, I.,
  {Zakamska}, N., {Anderson}, S., {Brinkmann}, J., {Bahcall}, N.~A., {Lamb},
  D.~Q., {Okamura}, S., {Szalay}, A., \& {York}, D.~G. 2003, \aj, 125, 1649

\bibitem[{{Ferland} {et~al.}(1998){Ferland}, {Korista}, {Verner}, {Ferguson},
  {Kingdon}, \& {Verner}}]{Cloudy98}
{Ferland}, G.~J., {Korista}, K.~T., {Verner}, D.~A., {Ferguson}, J.~W.,
  {Kingdon}, J.~B., \& {Verner}, E.~M. 1998, \pasp, 110, 761

\bibitem[{{Ferrarese} \& {Merritt}(2000)}]{FM00}
{Ferrarese}, L., \& {Merritt}, D. 2000, \apjl, 539, L9

\bibitem[{{Fukuda} {et~al.}(2000){Fukuda}, {Habe}, \& {Wada}}]{Fukuda00}
{Fukuda}, H., {Habe}, A., \& {Wada}, K. 2000, \apj, 529, 109

\bibitem[{{Gebhardt} {et~al.}(2000){Gebhardt}, {Bender}, {Bower}, {Dressler},
  {Faber}, {Filippenko}, {Green}, {Grillmair}, {Ho}, {Kormendy}, {Lauer},
  {Magorrian}, {Pinkney}, {Richstone}, \& {Tremaine}}]{Geb00}
{Gebhardt}, K., {Bender}, R., {Bower}, G., {Dressler}, A., {Faber}, S.~M.,
  {Filippenko}, A.~V., {Green}, R., {Grillmair}, C., {Ho}, L.~C., {Kormendy},
  J., {Lauer}, T.~R., {Magorrian}, J., {Pinkney}, J., {Richstone}, D., \&
  {Tremaine}, S. 2000, \apjl, 539, L13

\bibitem[{{Gnedin} {et~al.}(2004){Gnedin}, {Kravtsov}, {Klypin}, \&
  {Nagai}}]{Gnedinetal04}
{Gnedin}, O.~Y., {Kravtsov}, A.~V., {Klypin}, A.~A., \& {Nagai}, D. 2004, \apj,
  616, 16

\bibitem[{{Goldreich} \& {Lynden-Bell}(1965)}]{GoldLB65}
{Goldreich}, P., \& {Lynden-Bell}, D. 1965, \mnras, 130, 97

\bibitem[{{Kauffmann} \& {Haehnelt}(2000)}]{KaufHae00}
{Kauffmann}, G., \& {Haehnelt}, M. 2000, \mnras, 311, 576

\bibitem[{{Kennicutt}(1998)}]{Kennicutt98}
{Kennicutt}, Jr., R.~C. 1998, \apj, 498, 541

\bibitem[{{Klypin} {et~al.}(2001){Klypin}, {Kravtsov}, {Bullock}, \&
  {Primack}}]{Klypinetal01}
{Klypin}, A., {Kravtsov}, A.~V., {Bullock}, J.~S., \& {Primack}, J.~R. 2001,
  \apj, 554, 903

\bibitem[{{Konigl} \& {Kartje}(1994)}]{KonigKartje94}
{Konigl}, A., \& {Kartje}, J.~F. 1994, \apj, 434, 446

\bibitem[{{Kormendy} \& {Kennicutt}(2004)}]{KormKenn04}
{Kormendy}, J., \& {Kennicutt}, Jr., R.~C. 2004, \araa, 42, 603

\bibitem[{{Kravtsov}(1999)}]{KravtsovPhD}
{Kravtsov}, A.~V. 1999, PhD thesis, AA(NEW MEXICO STATE UNIVERSITY)

\bibitem[{{Kravtsov}(2003)}]{Kravtsov03}
---. 2003, \apjl, 590, L1

\bibitem[{{Kravtsov} {et~al.}(2002){Kravtsov}, {Klypin}, \&
  {Hoffman}}]{Kravtsovetal02}
{Kravtsov}, A.~V., {Klypin}, A., \& {Hoffman}, Y. 2002, \apj, 571, 563

\bibitem[{{Kravtsov} {et~al.}(1997){Kravtsov}, {Klypin}, \&
  {Khokhlov}}]{Kravtsovetal97}
{Kravtsov}, A.~V., {Klypin}, A.~A., \& {Khokhlov}, A.~M. 1997, \apjs, 111, 73

\bibitem[{{Kritsuk} {et~al.}(2007){Kritsuk}, {Norman}, {Padoan}, \&
  {Wagner}}]{Kritsuketal07}
{Kritsuk}, A.~G., {Norman}, M.~L., {Padoan}, P., \& {Wagner}, R. 2007, \apj,
  665, 416

\bibitem[{{Li} {et~al.}(2007){Li}, {Hernquist}, {Robertson}, {Cox}, {Hopkins},
  {Springel}, {Gao}, {Di Matteo}, {Zentner}, {Jenkins}, \&
  {Yoshida}}]{Lietal07}
{Li}, Y., {Hernquist}, L., {Robertson}, B., {Cox}, T.~J., {Hopkins}, P.~F.,
  {Springel}, V., {Gao}, L., {Di Matteo}, T., {Zentner}, A.~R., {Jenkins}, A.,
  \& {Yoshida}, N. 2007, \apj, 665, 187

\bibitem[{{Mac Low} \& {Klessen}(2004)}]{MacLowKlessen04}
{Mac Low}, M.-M., \& {Klessen}, R.~S. 2004, Reviews of Modern Physics, 76, 125

\bibitem[{{Machacek} {et~al.}(2001){Machacek}, {Bryan}, \& {Abel}}]{Machacek01}
{Machacek}, M.~E., {Bryan}, G.~L., \& {Abel}, T. 2001, \apj, 548, 509

\bibitem[{{Magorrian} {et~al.}(1998){Magorrian}, {Tremaine}, {Richstone},
  {Bender}, {Bower}, {Dressler}, {Faber}, {Gebhardt}, {Green}, {Grillmair},
  {Kormendy}, \& {Lauer}}]{Magorrian98}
{Magorrian}, J., {Tremaine}, S., {Richstone}, D., {Bender}, R., {Bower}, G.,
  {Dressler}, A., {Faber}, S.~M., {Gebhardt}, K., {Green}, R., {Grillmair}, C.,
  {Kormendy}, J., \& {Lauer}, T. 1998, \aj, 115, 2285

\bibitem[{{Malbon} {et~al.}(2006){Malbon}, {Baugh}, {Frenk}, \&
  {Lacey}}]{Malbonetal06}
{Malbon}, R.~K., {Baugh}, C.~M., {Frenk}, C.~S., \& {Lacey}, C.~G. 2006, ArXiv
  Astrophysics e-prints

\bibitem[{{Mayer} {et~al.}(2007){Mayer}, {Kazantzidis}, {Madau}, {Colpi},
  {Quinn}, \& {Wadsley}}]{Mayeretal07}
{Mayer}, L., {Kazantzidis}, S., {Madau}, P., {Colpi}, M., {Quinn}, T., \&
  {Wadsley}, J. 2007, Science, 316, 1874

\bibitem[{{McKee} \& {Ostriker}(2007)}]{McKeeOst07}
{McKee}, C.~F., \& {Ostriker}, E.~C. 2007, \araa, 45, 565

\bibitem[{{Modjaz} {et~al.}(2005){Modjaz}, {Moran}, {Greenhill}, \&
  {Kondratko}}]{Modjazetal05}
{Modjaz}, M., {Moran}, J.~M., {Greenhill}, L.~J., \& {Kondratko}, P.~T. 2005,
  in Astronomical Society of the Pacific Conference Series, Vol. 340, Future
  Directions in High Resolution Astronomy, ed. J.~{Romney} \& M.~{Reid}, 192--+

\bibitem[{{Navarro} {et~al.}(1997){Navarro}, {Frenk}, \& {White}}]{NFW97}
{Navarro}, J.~F., {Frenk}, C.~S., \& {White}, S.~D.~M. 1997, \apj, 490, 493

\bibitem[{{Nayakshin} \& {King}(2007)}]{NayakshinKing07}
{Nayakshin}, S., \& {King}, A. 2007, ArXiv e-prints, 705

\bibitem[{{Noguchi}(1988)}]{Noguchi88}
{Noguchi}, M. 1988, \aap, 203, 259

\bibitem[{{Ossenkopf} \& {Henning}(1994)}]{Ossenkopf94}
{Ossenkopf}, V., \& {Henning}, T. 1994, \aap, 291, 943

\bibitem[{{Ostriker} \& {Peebles}(1973)}]{OP73}
{Ostriker}, J.~P., \& {Peebles}, P.~J.~E. 1973, \apj, 186, 467

\bibitem[{{Passot} \& {V{\'a}zquez-Semadeni}(1998)}]{PassotVaz98}
{Passot}, T., \& {V{\'a}zquez-Semadeni}, E. 1998, \pre, 58, 4501

\bibitem[{{Polyachenko} {et~al.}(1997){Polyachenko}, {Polyachenko}, \&
  {Strel'Nikov}}]{Polyachenko}
{Polyachenko}, V.~L., {Polyachenko}, E.~V., \& {Strel'Nikov}, A.~V. 1997,
  Astronomy Letters, 23, 483

\bibitem[{{Pringle}(1981)}]{Pringle81}
{Pringle}, J.~E. 1981, \araa, 19, 137

\bibitem[{{Regan} \& {Teuben}(2004)}]{RegTeu04}
{Regan}, M.~W., \& {Teuben}, P.~J. 2004, \apj, 600, 595

\bibitem[{{Rice} {et~al.}(2005){Rice}, {Lodato}, \& {Armitage}}]{Riceetal05}
{Rice}, W.~K.~M., {Lodato}, G., \& {Armitage}, P.~J. 2005, \mnras, 364, L56

\bibitem[{{Ripamonti} \& {Abel}(2004)}]{RipAbel04}
{Ripamonti}, E., \& {Abel}, T. 2004, \mnras, 348, 1019

\bibitem[{{Roberts} {et~al.}(1979){Roberts}, {Huntley}, \& {van
  Albada}}]{Robertsetal79}
{Roberts}, Jr., W.~W., {Huntley}, J.~M., \& {van Albada}, G.~D. 1979, \apj,
  233, 67

\bibitem[{{Shlosman} \& {Begelman}(1989)}]{ShlosBeg89}
{Shlosman}, I., \& {Begelman}, M.~C. 1989, \apj, 341, 685

\bibitem[{{Shlosman} {et~al.}(1990){Shlosman}, {Begelman}, \&
  {Frank}}]{ShlosRev90}
{Shlosman}, I., {Begelman}, M.~C., \& {Frank}, J. 1990, \nat, 345, 679

\bibitem[{{Shlosman} {et~al.}(1989){Shlosman}, {Frank}, \&
  {Begelman}}]{barswinbars}
{Shlosman}, I., {Frank}, J., \& {Begelman}, M.~C. 1989, \nat, 338, 45

\bibitem[{{Sijacki} {et~al.}(2007){Sijacki}, {Springel}, {Di Matteo}, \&
  {Hernquist}}]{Sijacki07}
{Sijacki}, D., {Springel}, V., {Di Matteo}, T., \& {Hernquist}, L. 2007,
  \mnras, 380, 877

\bibitem[{{Simkin} {et~al.}(1980){Simkin}, {Su}, \& {Schwarz}}]{Simkinetal80}
{Simkin}, S.~M., {Su}, H.~J., \& {Schwarz}, M.~P. 1980, \apj, 237, 404

\bibitem[{{Springel} {et~al.}(2005){Springel}, {Di Matteo}, \&
  {Hernquist}}]{Springel05}
{Springel}, V., {Di Matteo}, T., \& {Hernquist}, L. 2005, \mnras, 361, 776

\bibitem[{{Toomre}(1964)}]{Toomre64}
{Toomre}, A. 1964, \apj, 139, 1217

\bibitem[{{Tremaine} {et~al.}(2002){Tremaine}, {Gebhardt}, {Bender}, {Bower},
  {Dressler}, {Faber}, {Filippenko}, {Green}, {Grillmair}, {Ho}, {Kormendy},
  {Lauer}, {Magorrian}, {Pinkney}, \& {Richstone}}]{Trem02}
{Tremaine}, S., {Gebhardt}, K., {Bender}, R., {Bower}, G., {Dressler}, A.,
  {Faber}, S.~M., {Filippenko}, A.~V., {Green}, R., {Grillmair}, C., {Ho},
  L.~C., {Kormendy}, J., {Lauer}, T.~R., {Magorrian}, J., {Pinkney}, J., \&
  {Richstone}, D. 2002, \apj, 574, 740

\bibitem[{{Truelove} {et~al.}(1997){Truelove}, {Klein}, {McKee}, {Holliman},
  {Howell}, \& {Greenough}}]{Truelove97}
{Truelove}, J.~K., {Klein}, R.~I., {McKee}, C.~F., {Holliman}, II, J.~H.,
  {Howell}, L.~H., \& {Greenough}, J.~A. 1997, \apjl, 489, L179+

\bibitem[{{Vazquez-Semadeni}(1994)}]{Vazquez94}
{Vazquez-Semadeni}, E. 1994, \apj, 423, 681

\bibitem[{{Vlemmings} {et~al.}(2007){Vlemmings}, {Bignall}, \&
  {Diamond}}]{Vlemmetal07}
{Vlemmings}, W.~H.~T., {Bignall}, H.~E., \& {Diamond}, P.~J. 2007, \apj, 656,
  198

\bibitem[{{Volonteri} \& {Rees}(2005)}]{VolRees05}
{Volonteri}, M., \& {Rees}, M.~J. 2005, \apj, 633, 624

\bibitem[{{Volonteri} \& {Rees}(2006)}]{VolRees06}
---. 2006, \apj, 650, 669

\bibitem[{{Wada}(2001)}]{Wada01}
{Wada}, K. 2001, \apjl, 559, L41

\bibitem[{{Wada} \& {Norman}(2001)}]{WadaNorman01}
{Wada}, K., \& {Norman}, C.~A. 2001, \apj, 547, 172

\bibitem[{{Wada} \& {Norman}(2007)}]{WadaNorman07}
---. 2007, \apj, 660, 276

\bibitem[{{Wong} \& {Blitz}(2002)}]{WongBlitz02}
{Wong}, T., \& {Blitz}, L. 2002, \apj, 569, 157

\bibitem[{{Wyse}(2004)}]{Wyse04}
{Wyse}, R.~F.~G. 2004, \apjl, 612, L17

\bibitem[{{Yoo} \& {Miralda-Escud{\'e}}(2004)}]{YooME04}
{Yoo}, J., \& {Miralda-Escud{\'e}}, J. 2004, \apjl, 614, L25

\bibitem[{{Young} {et~al.}(1996){Young}, {Allen}, {Kenney}, {Lesser}, \&
  {Rownd}}]{Youngetal96}
{Young}, J.~S., {Allen}, L., {Kenney}, J.~D.~P., {Lesser}, A., \& {Rownd}, B.
  1996, \aj, 112, 1903

\bibitem[{{Zweibel} \& {Heiles}(1997)}]{ZweibHeil97}
{Zweibel}, E.~G., \& {Heiles}, C. 1997, \nat, 385, 131

\end{thebibliography}


  \appendix

  \section{Measuring the Viscosity}
  \label{sec:mvisc}

  Here we address the validity of our estimate of the turbulent
  kinematic velocity parameter $\nu$, which motivates the
  interpretation that turbulent transport of angular momentum
  maintains local stability and the quasi-steady state of the
  circumnuclear disk. Instead of a direct measurement of the
  viscosity, we have assumed that the viscosity depends on
  characteristic velocity and length scales, given by
  $\sigma_{\textrm{turb}}$ and $\lambda_{\textrm{fast,r}}$,
  respectively. The accuracy of this estimate can be tested by
  measuring the transport of gas due to turbulent motions.

  The ART code does not follow individual parcels of gas across cell
  boundaries, but rather advected quantities describing the
  gas. Therefore, in order to follow the turbulent motions of the gas,
  we introduce a passive scalar into the simulation which has a value
  of unity inside a spherical region centered on the black hole
  particle, and a value of zero outside this region, effectively
  ``painting'' the gas. As the simulation evolves, turbulent motions
  of the gas cause the profile of the scalar quantity to deviate from
  its original spherical form. The evolution of the profile depends on
  the turbulent kinematic viscosity, $\nu$, and in the early stages of
  diffusion, while the deviations from the initial profile are small,
  the profile approximately satisfies the linear diffusion equation
  for initial conditions,

  \begin{equation}
    P(r,t=0) = 
    \begin{cases} 
      1,  & \mbox{if }r<r_0, \\
      0,  & \mbox{if }r>r_0, 
    \end{cases}
  \end{equation}

  \noindent (where $r_0$ is the radius of the initial sphere) allowing
  a more direct calculation of the viscosity. The solution to the
  linear diffusion equation for the above initial conditions, is

  \begin{equation}\label{eq:prof}
    P(x,q) = \frac{1}{2}\left[ \textrm{erf}\left(\frac{x+1}{\sqrt{q}}\right) -
    \textrm{erf}\left(\frac{x-1}{\sqrt{q}}\right)\right] - \frac{q}{x \sqrt{\pi
    q}}\ e^{-(1+x^2)/q}\ \sinh\left(\frac{2x}{q}\right),
  \end{equation}

  \noindent where

  \begin{equation}
    x = \frac{r}{r_0}\ \ \&\ \ q = \frac{4 \nu t}{r_0^2}.
  \end{equation}

\begin{figure}[t] 
    \centering
    \epsscale{1.}
    \plotone{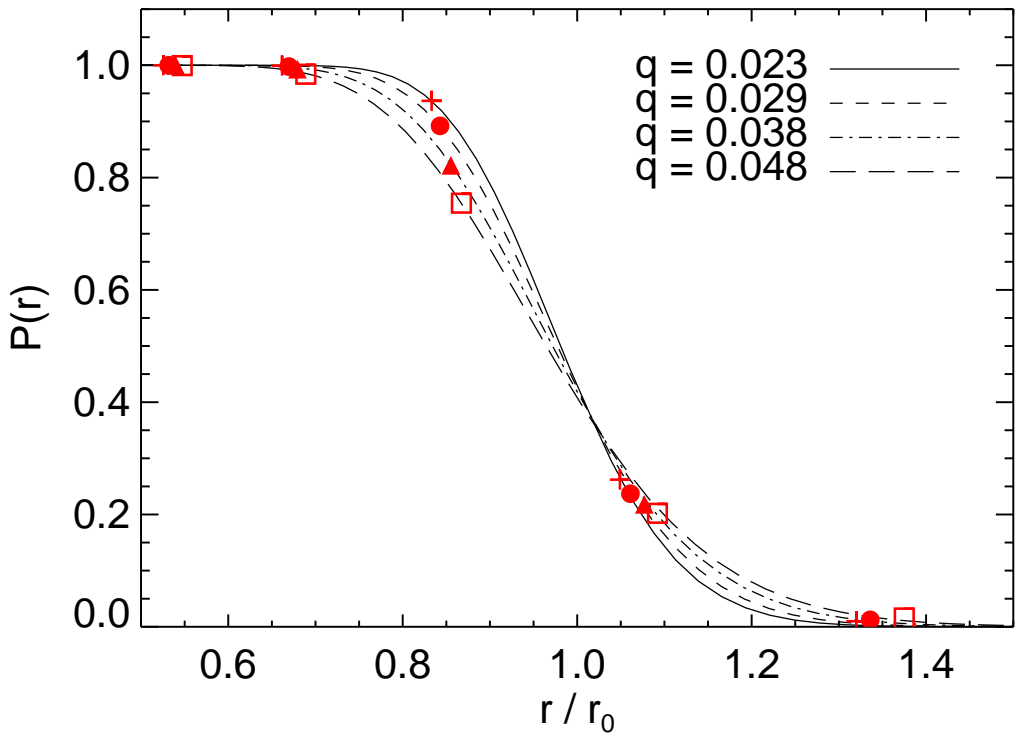}
    \caption{\label{fig:paint} Mass-weighted profiles of the passive
    scalar as a function of radius, and evolving over time from a $30$
    pc sphere. The symbols are measurements from the simulation, and
    the lines are the best fits to Equation \ref{eq:prof}.}
  \end{figure}

\begin{figure}[t] 
    \centering
    \epsscale{1.}
    \plotone{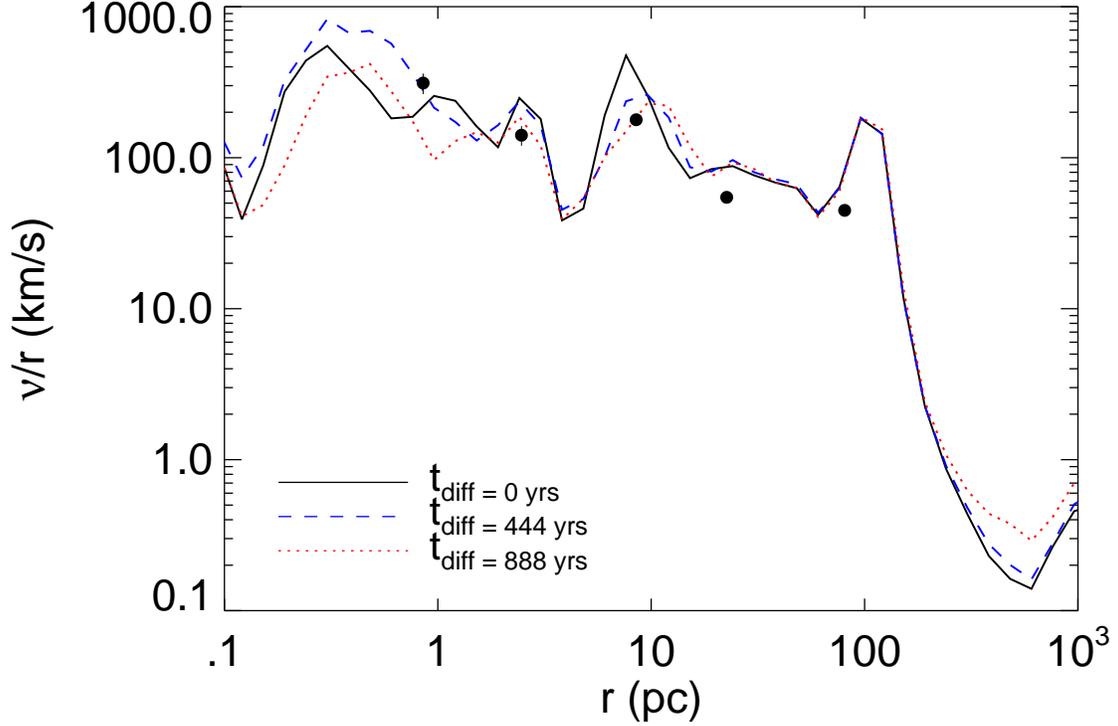}
    \caption{\label{fig:vpaint} Ratio of kinematic viscosity $\nu$, to
    radius $R$. The lines show estimates for $\nu/R$ given
    approximately by $\lambda_{\textrm{fast,r}}
    \sigma_{\textrm{turb}}$, for three different snapshots. The points
    are estimates from a second method measuring the advection of gas
    at five different radii in the simulation. The snapshots
    correspond to the profile fits at different times, so the $1$ pc
    point should be compared to the solid line, the next point to the
    dashed line, and the last three to the dotted line.}
  \end{figure}

  Figure \ref{fig:paint} shows samples of the profile evolving from a
  $30$ pc sphere for different times ($q$). The measured profiles
  agree well with the solution given in Equation
  \ref{eq:prof}. Starting from several painted regions at $1, 3, 10,
  30, \textrm{and}\ 100$ pc radii, we followed the evolution of the
  paint-weighted density profiles and measured the turbulent
  viscosity, $\nu$, as a function of radius. Figure \ref{fig:vpaint}
  shows the best-fit measurements of $\nu/r$ in relation to the
  estimate given by $\sigma_{\textrm{turb}} \lambda_{\textrm{fast,r}} /
  r$. The lines in Figure \ref{fig:vpaint} show individual snapshots
  of $\sigma_{\textrm{turb}} \lambda_{\textrm{fast,r}} / r$
  corresponding to the timescale over which the diffusion was followed
  (whereas Figure \ref{fig:visc} showed a time-averaged estimate).
  The best-fit measurements match the estimates for $\nu/r$ well, thus
  lending justification to the estimate for $\nu$ used in the analytic
  arguments of Section \ref{subsection:stab}.

  \section{Angular Momentum Conservation}
  \label{sec:con}

  Anomalous numerical transport of angular momentum is sometimes
  presented as a concern for adaptive mesh refinement simulations
  using interpolation schemes for calculating velocities on the
  mesh. We have tested conservation of angular momentum in the code by
  conducting a separate set of simulations of the collapse of an
  isothermal sphere (the details are included in the PhD thesis of the
  author Levine). The test was conducted for a mesh which refined
  along with the collapsing sphere, and for a pre-refined mesh, in
  order to test the conservation of angular momentum both as the mesh
  refines and as gas moves through the mesh. The sphere collapses into
  a thin disk ($h/r << 1$), which conserves angular momentum over
  several rotation periods.

\begin{figure}[t]
    \centering
    \epsscale{1.}
    \plotone{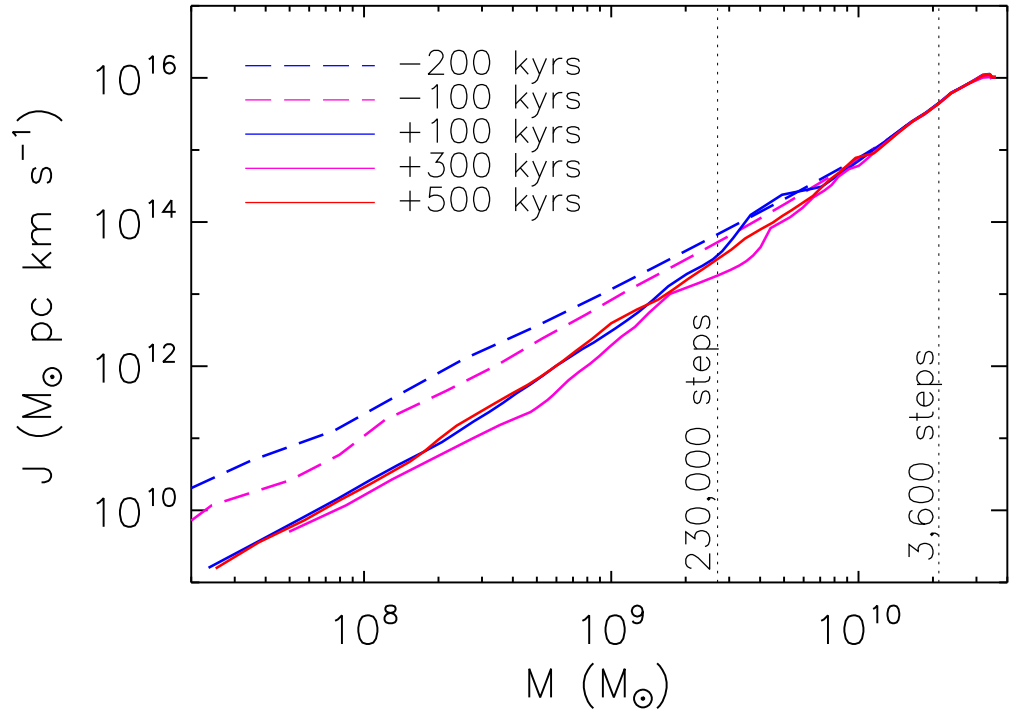}
    \caption{\label{fig:jvm} Angular momentum as a function of
    enclosed gas mass for several different times. The times are measured
    from the introduction of the black hole. The vertical dashed lines
    correspond to enclosed masses on scales of $\approx 10$ and $\approx
    1000$ pc. At each of these scales, the simulation has made
    approximately $230{,}000$ and $3{,}600$ steps, respectively
    (corresponding to the resolution at each scale), between $t=100$
    and $t=500$ kyrs.}
  \end{figure}

  In Figure \ref{fig:jvm} we demonstrate the conservation of angular
  momentum in the present simulation. As the maximum resolution of the
  simulation increases, the angular momentum profile initially
  evolves, as the viscous torque, $G$, described in Section
  \ref{subsection:ang} redistributes angular momentum within the
  disk. But after reaching the maximum, level $20$ resolution (the
  point at which we introduce the black hole particle into the
  simulation and let it evolve), the profile remains rather steady
  with time, because the disk has reached a quasi-stationary
  state. Figure \ref{fig:jvm} shows the angular momentum as a function
  of enclosed gas mass at several different times during a single
  zoom-in episode of the simulation. The figure shows the evolution of
  the angular momentum distribution from $200,000$ years before the
  introduction of the black hole particle, when the simulation has
  refined to level $11$, to $500,000$ years after the introduction of
  the black hole particle, when the simulation is fully refined, and
  has evolved for several hundred thousand years in a quasi-steady
  state.

  On the scale of the circumnuclear disk, the simulation has made
  several thousand time steps between $t=100$ kyrs and $t=500$ kyrs,
  meaning that the simulation has undergone significant evolution on
  this spatial scale. Numerical effects, accumulating with each time
  step, would cause deviations in the angular momentum profile. Over
  longer time scales, the slow inward transport of matter will alter
  the angular momentum profile. However, on the hundred thousand year
  time scale we expect the profile to remain rather steady, reflecting
  the quasi-stationary state of the disk, unless there is anomalous
  numerical transport of angular momentum. Figure \ref{fig:jvm} shows
  no systematic deviations in the angular momentum profile after the
  initial refinement and the corresponding re-distribution of the
  angular momentum, demonstrating that there is no sign of unphysical
  numerical angular momentum transport on the time scales of the
  simulation. On time scales much longer than those simulated in the
  present paper, however, these effects may become important.

\end{document}